\documentclass[]{interact}
\usepackage{enumitem}
\usepackage{epstopdf}
\usepackage[caption=false]{subfig}

\usepackage[numbers,sort&compress,merge]{natbib}
\bibpunct[, ]{[}{]}{,}{n}{,}{,}
\renewcommand\bibfont{\fontsize{10}{12}\selectfont}

\theoremstyle{plain}

\theoremstyle{definition}

\theoremstyle{remark}

\usepackage{color,soul}
\usepackage{xspace}
\usepackage{xcolor}

\usepackage{lmodern}

\begin{document}

\articletype{ARTICLE TEMPLATE}

\title{Monte Carlo Simulations of the Blinking-Checkers Model for Polyamorphic Fluids}

\author{
\name{Sergey V. Buldyrev\textsuperscript{a,b}\thanks{Corresponding Authors: S.~V. Buldyrev (buldyrev@yu.edu) and F. Caupin (frederic.caupin@univ-lyon1.fr).}, Thomas J. Longo\textsuperscript{c}, Fr\'ed\'eric Caupin\textsuperscript{d}, and Mikhail A. Anisimov\textsuperscript{c,e}}
\affil{\textsuperscript{a}Department of Physics, Yeshiva University, New York, New York 10033, United States; 
\textsuperscript{b}Department of Physics, Boston University, Boston, Massachusetts 02215, United States;
\textsuperscript{c}Institute for Physical Science and Technology, University of Maryland, College Park, Maryland 20742, United States; 
\textsuperscript{d}Institut Lumi\`ere Mati\`ere, Universit\'e Claude Bernard Lyon 1, CNRS, Institut Universitaire de France, F-69622 Villeurbanne, France; \textsuperscript{e}Department of Chemical and Biomolecular Engineering, University of Maryland, College Park, Maryland 20742, United States}
}

\maketitle

\begin{abstract}
The blinking-checkers model [F. Caupin and M. A. Anisimov, \textit{Phys. Rev. Lett}, \textbf{127},185701 (2021)] is a minimal lattice model which has demonstrated that, in the meanfield approximation, it can reproduce the phenomenon of fluid polyamorphism. This model is a binary lattice-gas, in which each site has three possible states: empty, occupied with particles of type 1, and occupied with particles of type 2. Additionally, the two types of particles may interconvert from one to another. Equilibrium interconversion imposes a constraint that makes this model thermodynamically equivalent to a single-component system. In this work, Monte-Carlo simulations of the blinking-checkers model are performed, demonstrating polyamorphic phase behavior. The locations of the liquid-liquid and liquid-gas critical points are found to be different from the meanfield predictions for this model with the same interaction parameters, as the phase behavior is significantly affected by critical fluctuations. Based on the computed values of the critical exponents of the order parameter, susceptibility, correlation length, and surface tension, we confirm that the blinking-checkers model, for both liquid-gas and liquid-liquid equilibria, belongs to the three-dimensional Ising class of critical-point universality.
\end{abstract}


\section{Introduction}
Typically, pure substances may be found with only one gaseous or liquid state, while their solid state may exist in various polymorphic crystalline states. The existence of multiple distinct liquid forms in a single-component substance is more unusual since liquids lack the long-range order common to crystals. Yet, the existence of multiple amorphous liquid states in a single component substance, a phenomenon known as ``liquid polyamorphism,''~\cite{Stanley_Liquid_2013,Anisimov_Polyamorphism_2018,Tanaka_Liquid_2020} has been observed or predicted in a wide variety of substances, such as superfluid helium~\cite{Vollhardt_He_1990,Schmitt_He_2015}, high-pressure hydrogen~\cite{Ohta_H_2015,Zaghoo_H_2016,McWilliams_H_2016,Norman_Review_2021,Fried_Hydrogen_2022}, high-density  sulfur~\cite{Henry_Sulfur_2020,Shumovskyi_Sulfur_2022}, phosphorous~\cite{Katayama_Phos_2000,Katayama_Phos_2004}, carbon~\cite{Glosli_Liquid_1999}, silicon~\cite{Sastry_Silicon_2003,Beye_Silicon_2010,Vasisht_Solicon_2011,Sciorino_Silicon_2011}, selenium and tellurium~\cite{Tsuchiya_SeTe_1982,Brazhkin_SeTe_1999}, cerium~\cite{Cadien_Ce_2013}, and in various oxides~\cite{Tanaka_Liquid_2020} - \textit{e.g.} silica~\cite{Saika_Silica_2000,Lascaris_Silica_2014}. It has also been hypothesized in deeply supercooled liquid water~\cite{Angell_TwoState_1971,Angell_Amorphous_2004,Stanley_Liquid_2013,Anisimov_Polyamorphism_2018,Tanaka_Liquid_2020,Poole_Water_1992,Debenedetti_Water_1998,Holten_Water_2012,Holten_Water_2014,Gallo_Water_2016,Biddle_Water_2017,Caupin_Thermodynamics_2019,Duska_Water_2020}.

A substance may be found to be polyamorphic by experimentally or computationally detecting a liquid-liquid phase transition (LLPT), which may terminate at a liquid-liquid critical point (LLCP) or a triple point. For example, liquid polyamorphism, via the existence of two alternative supramolecular structures, may explain the remarkable anomalies in the thermodynamic properties of supercooled water, namely a maximum in the temperature dependence of its density and its isothermal compressibility~\cite{Holten_Compressibility_2017,Kim_Maxima_2017}, a maximum in the isobaric heat capacity~\cite{Pathak_Enhancement_2021}, and an inflection point in its surface tension~\cite{Hruby_Surface_2014,Vins_Surface_2015,Vins_Surface_2017,Vins_Surface_2020}. Simulations of water-like models~\cite{Poole_Water_1992,Holten_Water_2012,Holten_Water_2014,Gonzalez_Comprehensive_2016,Palmer_Anomalous_2018,Singh_Thermodynamic_2019,Debenedetti_Water_1998,Biddle_Water_2017,Debenedetti_Second_2020}, potentially supported by experiment~\cite{Mishima_Decompression_1998,Kim_WaterExp_2020}, have demonstrated the hypothesized LLPT in supercooled water.  For instance, TIP4P/2005, a rigid model for water introduced by Abascal and Vega in 2005~\cite{Abascal_General_2005}, is arguably one of the most realistic interaction potentials for water~\cite{Vega_Simulating_2011}. It exhibits the same thermodynamic anomalies as real water, and it predicts more anomalies in the metastable regions of the phase diagram~\cite{Gonzalez_Comprehensive_2016}. In particular, in the negative pressure region, TIP4P/2005 predictions compare well with experiments~\cite{Pallares_Anomalies_2014,Pallares_Equation_2016,Holten_Compressibility_2017}. Recently, state-of-the-art simulations (with histogram reweighting and large-system scattering calculations) proved the existence of a LLCP in TIP4P/2005 water, consistent with the three-dimensional Ising universality class~\cite{Debenedetti_Second_2020}.

Liquid (or, more generally, fluid) polyamorphism ultimately originates from the complex interactions between molecules or supramolecular structures. However, it may be phenomenologically modeled through the reversible interconversion of two alternative molecular or supramolecular states \cite{Anisimov_Polyamorphism_2018,Caupin_Minimal_2021,MFT_PT_2021}. The application of this ``two-state'' approach to the variety of polyamorphic substances could be a useful phenomenology, or it may reflect the true microscopic origin of fluid polyamorphism. Indeed, there are a few substances, such as hydrogen, sulfur, phosphorous, and carbon, where the existence of alternative liquid or dense-fluid states is explicitly induced by a reversible chemical reaction: dimerization in hydrogen~\cite{Fried_Hydrogen_2022,Shumovsky2023} or polymerization in sulfur, phosphorus, and carbon~\cite{Shumovskyi_Sulfur_2022,Buldyrev2024}.

The interconversion of two-states significantly affects both the thermodynamics and dynamics of fluid mixtures. In particular, fast molecular interconversion imposes an additional thermodynamic constraint, the chemical-reaction equilibrium condition, which reduces the number of thermodynamic degrees of freedom. In this case, the concentration of a binary mixture is no longer an independent variable, being a function of temperature and pressure. Consequently, a binary mixture with interconversion of species follows the Gibbs phase rule for a single-component substance. In the absence of interconversion, the binary mixture exhibits a critical point for each concentration. Collectively, these critical points make up a critical locus. The interconversion of species selects a single path through the planes of phase coexistence at fixed concentration, crossing the critical locus of the binary mixture at a unique point. Moreover, this path could cross the critical line, the line of triple points, or any other unique line on the phase diagram more than once. The resulting phase diagram can be viewed as the phase diagram of a polyamorphic single-component fluid with multiple fluid-fluid critical points.

Interestingly, a phenomenological meanfield model incorporating interconversion between two states was shown to be able to reproduce quantitatively thermodynamic properties of real~\cite{Caupin_Thermodynamics_2019} and TIP4P/2005~\cite{Biddle_Water_2017} water. A simpler model, based on the two-state approach, that is still capable of qualitatively describing the anomalies in the thermodynamic properties of supercooled water (and systems exhibiting similar anomalies) is the ``blinking-checkers lattice model''~\cite{Caupin_Minimal_2021,Longo_Interfacial_2023}. This is a compressible binary-lattice model, thermodynamically equivalent to the mixture of two lattice gases, in which ``blue'' (state 1) and ``red'' (state 2) species interact, while interconversion of species is implemented by allowing the particles to change their ``color'' by tuning the energy and/or entropy of interconversion. Previous lattice models were able to reproduce the anomalous behavior of water through complex interactions between particles~\cite{Sastry_SingularityFree_1996,Hruby_Twostructure_2004,Ciach_Model_2008}, while the blinking-checkers model represents a minimalistic microscopic approach. Additionally, the blinking-checkers model may be phenomenologically applied to describe the structural interconversion of any two states, like that predicted for supercooled water. 

Building on previous studies of the blinking-checkers model~\cite{Caupin_Minimal_2021,Longo_Interfacial_2023}, in this work, we account for critical fluctuations of concentration or density by conducting Monte Carlo (MC) simulations of the model. The rest of this manuscript is outlined as follows. In section 2, the application of the MC method to the blinking-checkers model is described. A method for calculating the pressure of a MC system is reported. In section 3, the behavior of the blinking-checkers model in the vicinity of the liquid-gas critical point (LGCP) and LLCP are investigated through the coexistence, isothermal compressibility (susceptibility), correlation length, and surface tension. It is demonstrated that both critical points belong to the three-dimensional Ising-model class of critical-point universality. 

\section{Monte Carlo Simulation Technique}

MC simulations of the blinking-checkers model were performed on a $\ell_x\times\ell_y\times\ell_z $ simple cubic lattice with $\ell_x\ell_y\ell_z = n$ total sites and periodic boundary conditions. Simulations of coexistence were performed in a rectangular box with sides, $\ell_x=\ell_y=\ell_z/2 \equiv \ell$, while all other simulations were performed in a cubic box with length $\ell_x=\ell_y=\ell_z\equiv\ell$. Systems of size $\ell= 32$ to $\ell = 256$ were investigated. Each lattice site, $i$, may be in one of three possible states: $s_i=0$ (empty), $s_i=1$ (particle of type 1), and $s_i=2$ (particle of type 2). The number of empty sites, $n_0$, is fixed, while the number of sites filled with particles of types 1 and 2 (given by $n_1$ and $n_2$, respectively) are varied through an interconversion reaction. The number of sites in each state are related to the total number of sites as $n_0+n_1+n_2\equiv n$. The overall density and molecular fractions are defined as in Ref.~\cite{Caupin_Minimal_2021}, as $\bar{\rho}=(n_1+n_2)/n$, $x_1=n_1/(n_1+n_2)$, and $x_2=n_2/(n_1+n_2)$. 

We use a nearest-neighbor approximation, in which each lattice site $i$ has $z=6$ nearest neighbors, $j$, belonging to the nearest neighborhood, $\mathcal{L}_i$, of site $i$ which may interact with site $i$, such that the potential energy of site $i$ is given by 
\begin{equation}\label{Eqn_IntEn_perSite}
    u_i=\sum_{j\in\mathcal{L}_i}\epsilon(s_{i},s_{j})
\end{equation}
where $\epsilon(s_i,s_j)=2\omega_{ij}/z$ is a symmetric matrix, such that $\omega_{ij}=\omega_{ji}$. Interactions with an empty site, $s_i=0$, are assumed to be zero. In this work, following the recent meanfield study of this model~\cite{Longo_Interfacial_2023}, the case of $\omega_{11}=2.0$, $\omega_{12}=\omega_{21}=1.4$, and $\omega_{22}=2.5$ is considered.

\subsection{Interconversion ``Flipping'' and Diffusion ``Swapping'' Steps}

At each MC step, a diffusion ``swapping'' (Kawasaki) step~\cite{kawasaki_diffusion_1966} or an interconversion ``flipping'' (Glauber) step~\cite{glauber_timedependent_1963} is attempted (see details in Ref.~\cite{Shum_Phase_2021}). For either step, the change in the free energy is computed as $\Delta F = \Delta U -T\Delta S$, where $\Delta U$ is the change in the total potential energy caused by the attempted swap or flip. The potential energy of the system is given by the application of Eq.~(\ref{Eqn_IntEn_perSite}) to all sites as
\begin{equation}\label{eq:UMC}
    U=\frac{1}{2}\sum_i^n\sum_{j\in\mathcal{L}_i}\epsilon(s_i,s_j)+\tilde{e} n_2
\end{equation}
where $\tilde{e}$ is the internal energy associated with state, $s_i=2$.

\begin{figure}[t]
    \centering
    \includegraphics[width=0.49\linewidth]{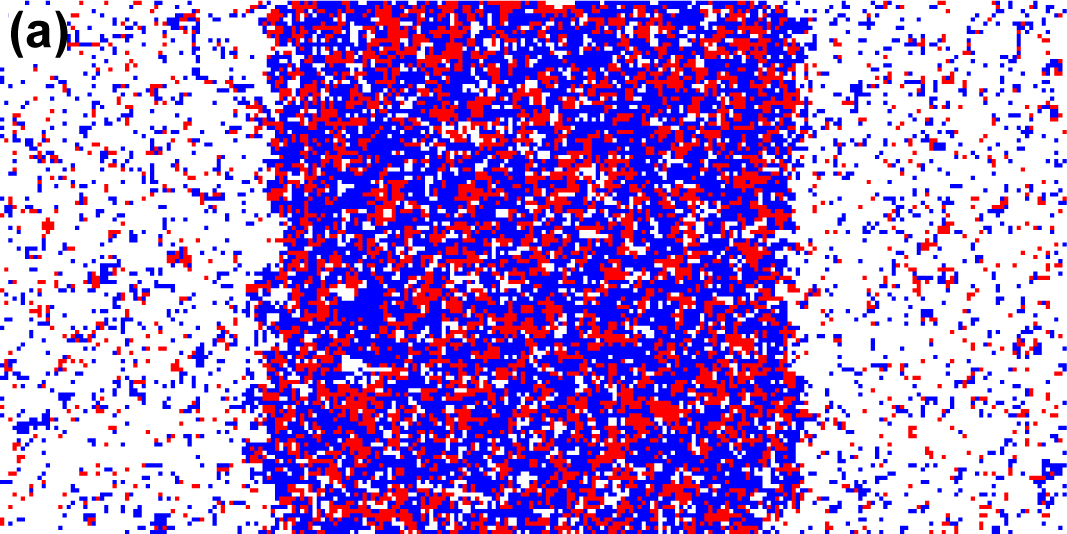}
    \includegraphics[width=0.49\linewidth]{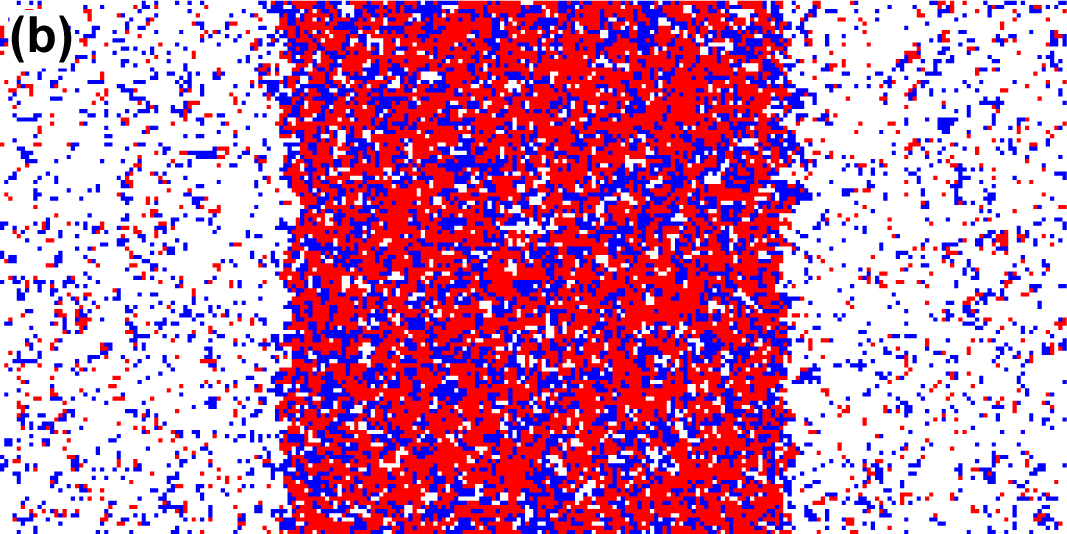}
    \includegraphics[width=0.49\linewidth]{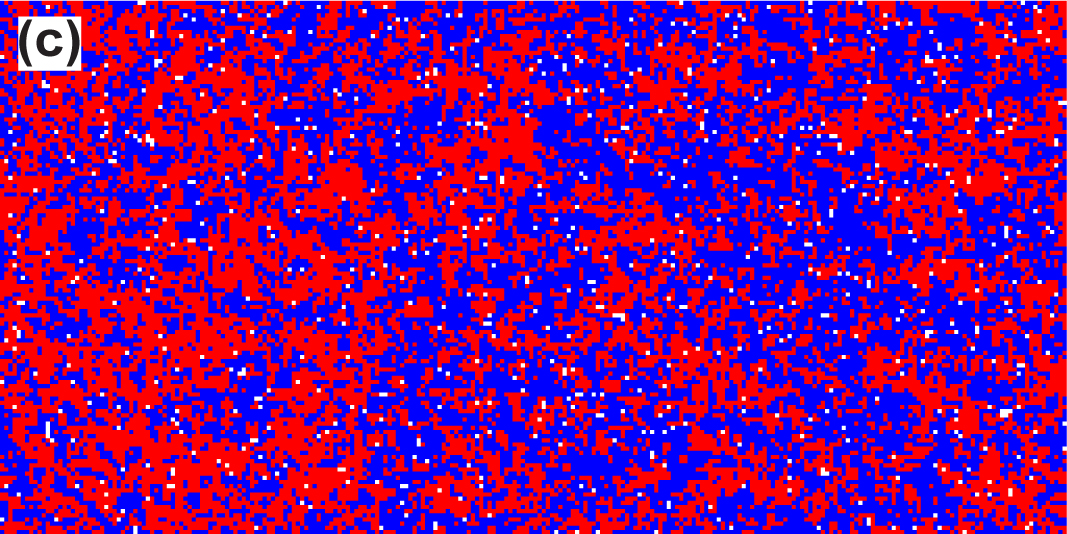}
    \includegraphics[width=0.49\linewidth]{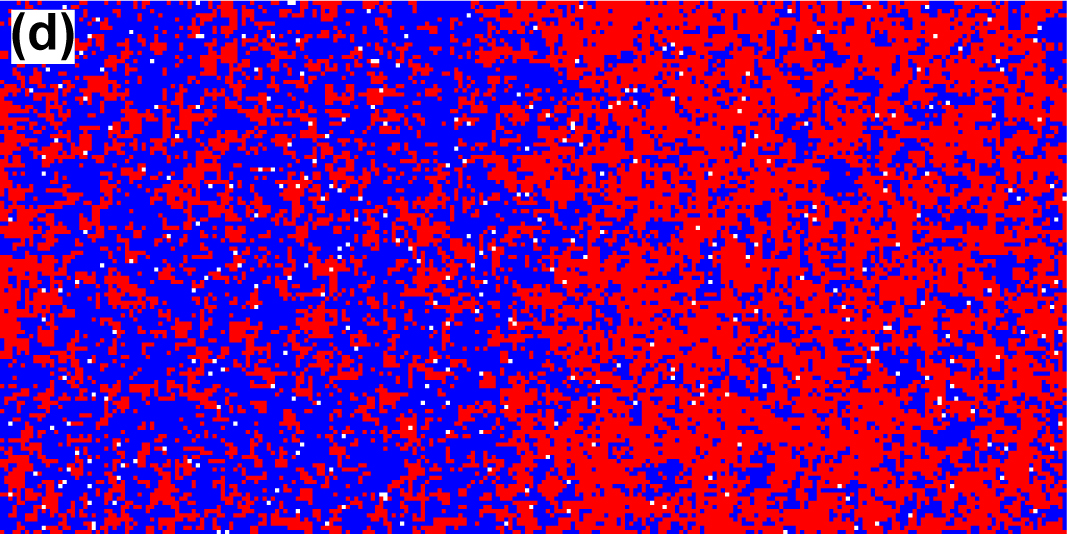}
    \caption{Two-dimensional slices of the MC simulations for the system with $\omega_{11} =2.0$, $\omega_{12}=1.4$, and $\omega_{22} = 2.5$, at four different temperatures and average densities, $\bar{\rho}$. The panels correspond to the following conditions: (a) $T=0.63$ and $\bar{\rho} = 0.5$, (b) $T = 0.64$ and $\bar{\rho} = 0.5$, (c) $T = 0.6269$ and $\bar{\rho}=0.980$, and (d) $T = 0.6260$ and $\bar{\rho}=0.990$. Liquid-gas equilibrium is depicted in (a,b), while (c) and (d) show the system near the LLCP temperature: just above (supercritical, (c)) and just below (liquid-liquid equilibrium, (d)). The blue cells correspond to atoms of type 1, while the red cells correspond to atoms of type 2.}
    \label{Fig_Snapshots}
\end{figure}

For any Kawasaki swap, the change in the entropy, $\Delta S=0$, while for a Glauber flip $\Delta S=\pm \tilde{s}$, where $\tilde{s}$ is the internal entropy of the reaction, such that ``$+$'' indicates when a state of this site changes from 1 to 2, while ``$-$'' indicates the opposite scenario. In accordance with the Metropolis criterion~\cite{metropolis_basic_1963}, the new state is accepted with probability $p=\exp(-\Delta F/T)$ for $\Delta F>0$, while for $\Delta F<0$, the new state is always accepted. In this work, just as in Ref.~\cite{Caupin_Minimal_2021}, the case $\tilde{e}=3$ and $\tilde{s}=4$ is considered. With use of these parameters, simulation snapshots for this system at different temperatures and densities are presented in Fig.~\ref{Fig_Snapshots}. 

Calculations of pressure are generally challenging in MC simulations. In the main text of this work, we employ the Widom-insertion method, detailed in Section~\ref{Sec_WidomMethod}. We also considered a more contemporary method, which utilizes ``ghost'' lattice sites and is discussed in Appendix~\ref{Appendix}.

\subsection{Widom-Insertion Method for Pressure Calculations}\label{Sec_WidomMethod}
The traditional method for calculating the pressure in MC simulations is known as the ``Widom-insertion'' method~\cite{Widom63}. In this work, this method is adjusted to apply to a system of several species. In the adjusted Widom-insertion method, the pressure is calculated by first attempting to insert a particle of type 1 in all $n$ sites one-by-one. If a site is occupied, the change in the potential energy due to insertion is $\Delta u_i=\infty$, while if the cell is empty, $\Delta u_{1,i}=\sum_{j\in \mathcal{L}_i}\epsilon(1,s_{j})$. The insertion parameter for a particle of type 1, is given by
\begin{equation}
    B_1=\frac{1}{n}\sum_{i=1}^n  e^{-\Delta u_{1,i}/k_\text{B}T}
\end{equation}
Analogously, $B_2$ is computed by the insertion of a particle of type 2. Note that for the case of a particle of type 2, there is no need to include the term with $\tilde{e}$, as in Eq.~(\ref{eq:UMC}), since this term is canceled in the final calculation of the pressure. 
The total excess chemical potential is defined as
\begin{equation}
   \mu_\text{ex}=-k_\text{B}T(x_1\ln B_1+x_2\ln B_2)
\end{equation}
and is averaged over time with an interval of one MC time unit ($n$ MC steps) after the system reaches equilibrium. This calculation must be repeated for all densities from $\rho'=0$ to the density of the system at a given state point $\rho'=\bar{\rho}$ with a sufficiently small step size, $\Delta \bar{\rho}$, at fixed $T$ and without interconversion, so that the mole fractions, $x_1$ and $x_2$ are fixed to the values obtained for the state point $(\bar{\rho}, T)$ with interconversion. Finally, the pressure is obtained from the thermodynamic relation  $(\partial P/\partial\bar{\rho})_T=\bar{\rho}(\partial\mu/\partial \bar{\rho})_T$ as 
\begin{equation}\label{Eq_Pressure_Widom}
    P=\bar{\rho} \left[k_\text{B} T +\mu_\text{ex}(\bar{\rho}) -\frac{1}{\bar{\rho}}\int_0^{\bar{\rho}} \mu_\text{ex}(\rho^\prime )d\rho^\prime\right]
\end{equation}
We note that the Widom-insertion method is a computationally expensive calculation. A contemporary (less expensive) calculation of the pressure is considered in Appendix~\ref{Appendix}. However, since it was found that the increase in the computational efficiency of this alternative method came at a cost to its accuracy, all further calculations of pressure presented in the main text were performed with the Widom-insertion method.

\section{Results and Discussion}
In this work, the liquid-gas and liquid-liquid phase coexistence, isothermal compressibility, correlation length, and interfacial tension for a system of varying sizes with $\omega_{11} = 2.0$, $\omega_{12} = 1.4$, $\omega_{22} = 2.5$, $\tilde{e}=3$, and $\tilde{s}=4$ was investigated. The following three subsections detail the results of the analysis of these properties.

\subsection{Liquid-Gas and Liquid-Liquid Coexistence}

\begin{table}[tp!]
\centering
\caption{Critical exponents of thermodynamic properties of an Ising-like system and those predicted by the meanfield approximation~\cite{Ch8_Hassan_Mesothermo_2010}.}
\label{Table_Critical_Exponents}
\begin{tabular}{lccc}
\toprule
Thermodynamic property & Critical exponent & 3$d$ Ising system & Meanfield value \\ \midrule
Heat capacity          & $\alpha$          & $0.110$           & 0               \\
Order parameter        & $\beta$           & $0.326$           & 0.5             \\
Susceptibility         & $\gamma$          & $1.240$           & 1               \\
Correlation length     & $\xi$             & $0.630$           & 0.5             \\
Correlation function   & $\eta$            & $0.033$           & 0               \\
Surface tension        & $\mu$             & $1.260$           & 1.5      \\ \bottomrule      
\end{tabular}
\end{table}

\begin{figure}[tp!]
    \centering
    \includegraphics[width=0.51\linewidth]{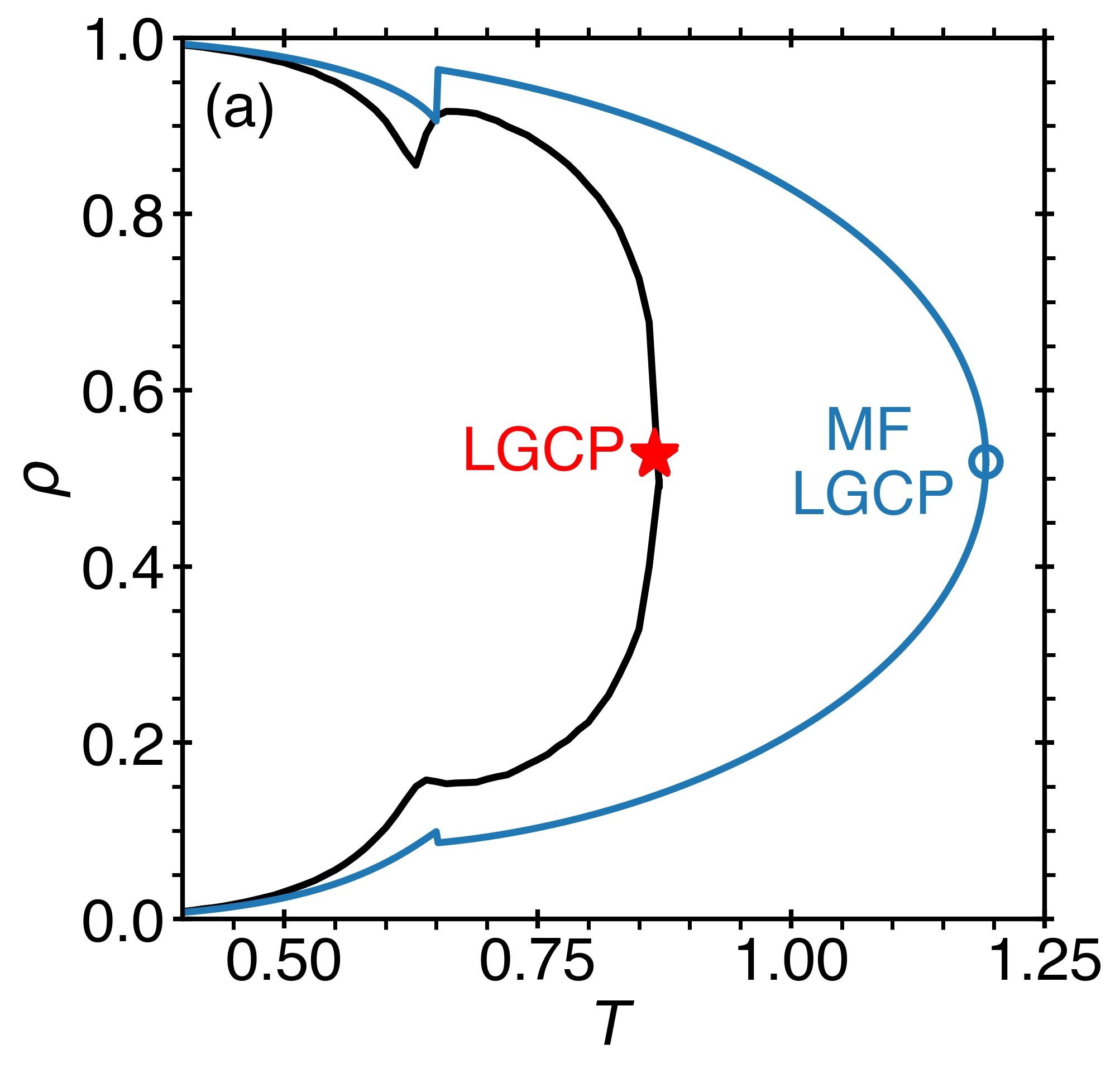}
    \includegraphics[width=0.47\linewidth]{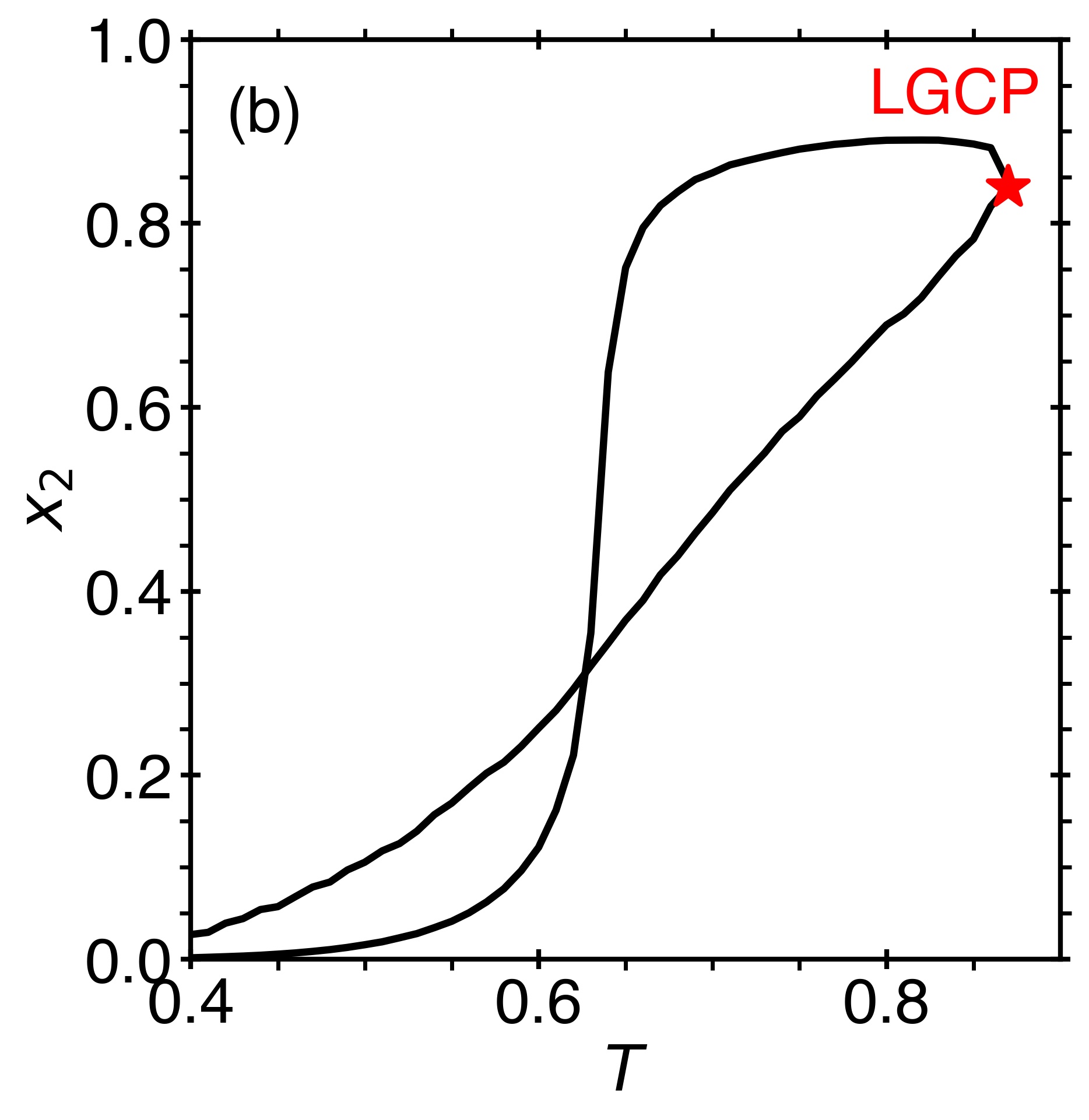}
    \includegraphics[width=0.49\linewidth]{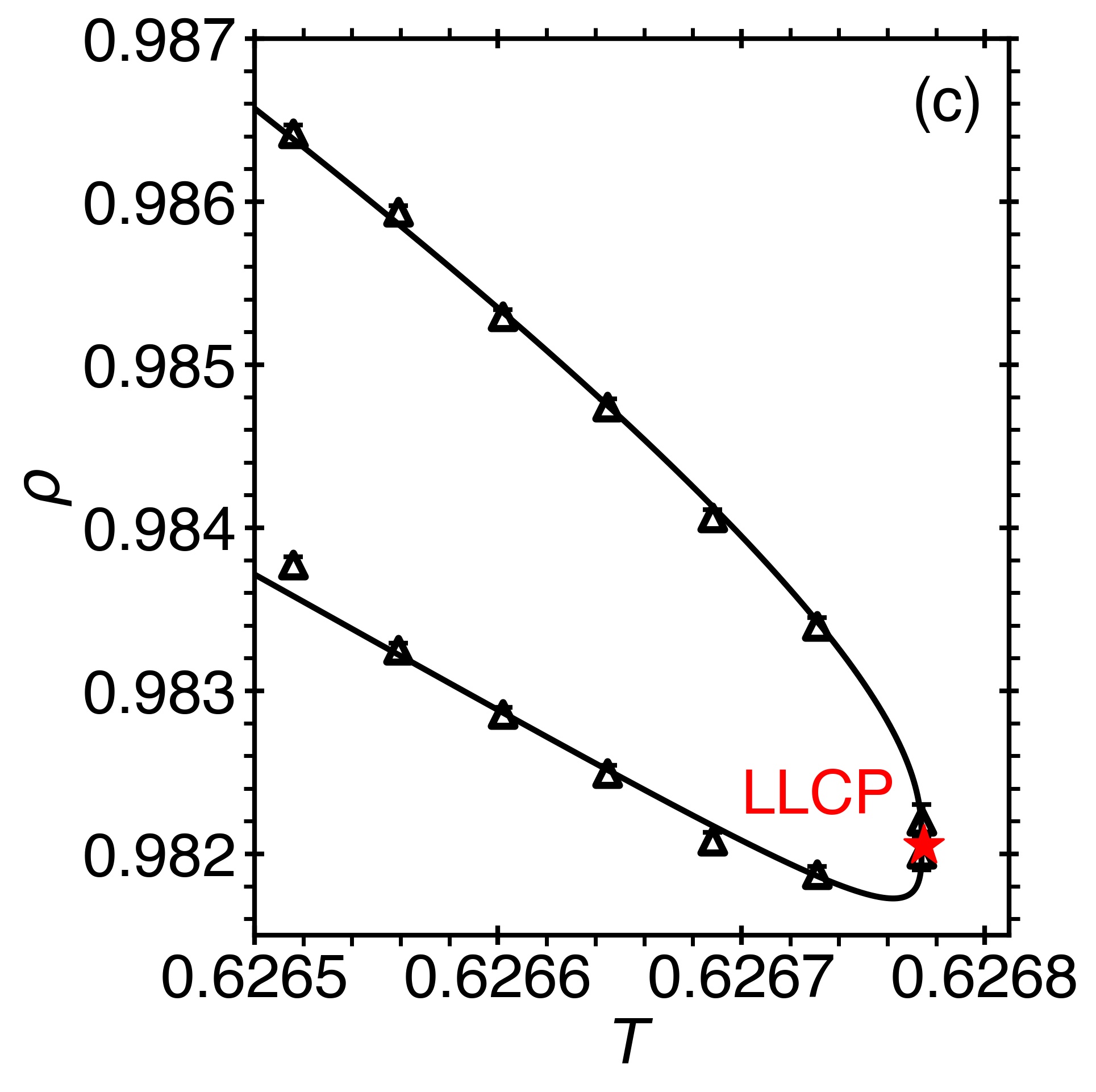}
    \includegraphics[width=0.49\linewidth]{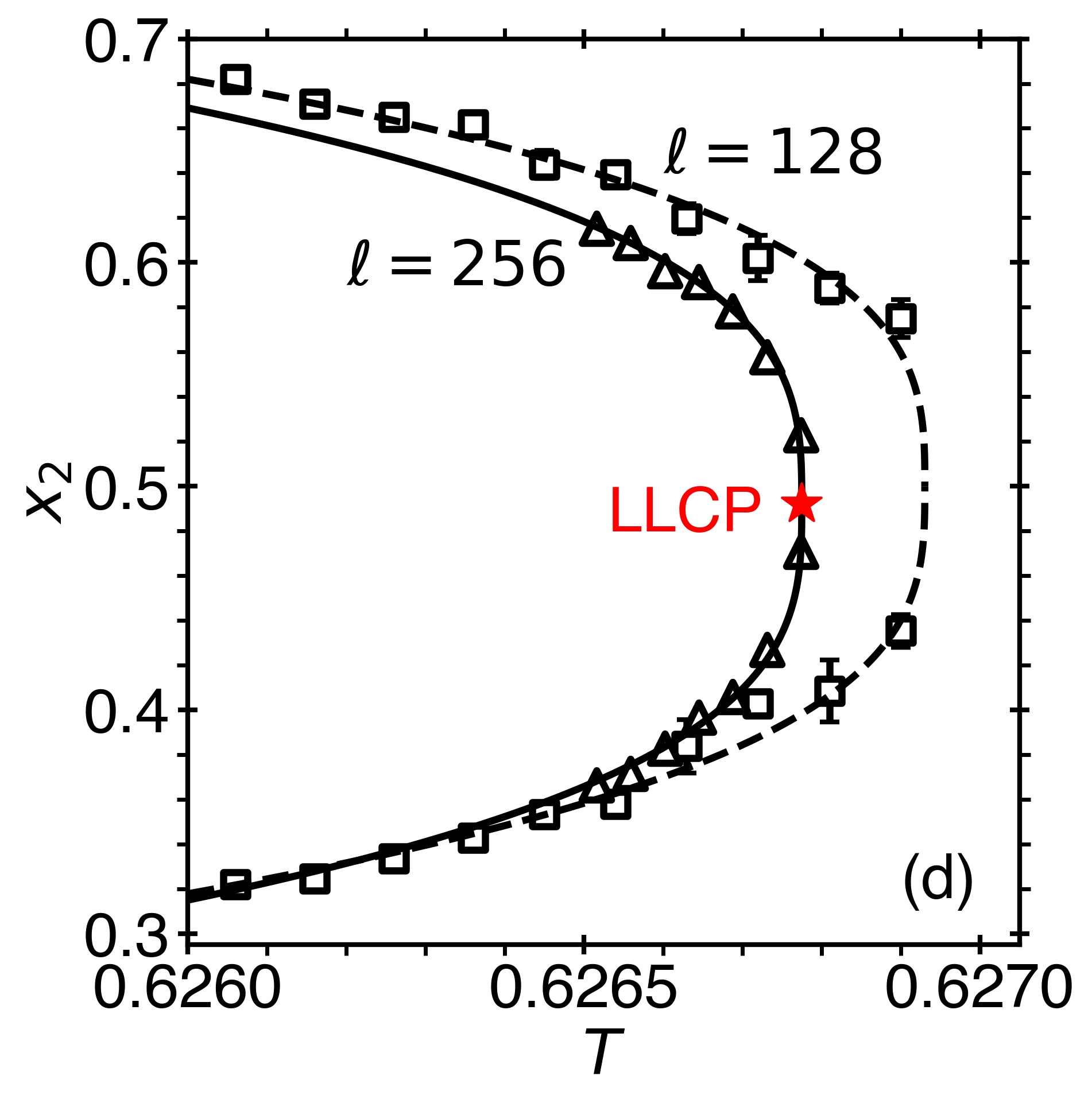}
    \caption{Liquid-gas (a,b) and liquid-liquid (c,d) coexistence curves, in which $\rho$ is a density of a coexisting phase and $x_2$ is the molecular fraction of state 2, for the system with $\omega=1.4$ and system size $\ell=256$, where $\ell_y=\ell_x=\ell_z/2\equiv \ell$, The LGCP and LLCP are indicated by the red stars. In (a), the meanfield prediction of the coexistence for a same system with the same parameters is shown in blue. The meanfield prediction of the LGCP (MF-LGCP) is indicated by the open blue circle. In (c,d), the solid curves represent the predictions of the scaling theory, given by Eqs.~(\ref{Eq_deltaRhoLL}) and (\ref{Eq_deltaXLL}), while in (d) the dashed curve represents a guideline for the data of size $\ell = 128$, highlighting the shift in the critical temperature to lower values for larger system sizes.}
    \label{Fig_Phase_Diagrams}
\end{figure}

MC simulations of coexistence were obtained in a box with sides, $\ell_x=\ell_y=\ell_z/2$. The liquid-gas (LG) and liquid-liquid (LL) coexistence curves are presented in Fig.~\ref{Fig_Phase_Diagrams}, where $\rho$ is the density of a coexisting phase and $x_2$ is the molecular fraction of state 2. Away from the LGCP and LLCP, the results of the MC simulations are in agreement with the predictions of the meanfield theory~\cite{Longo_Interfacial_2023}. As depicted in Fig.~\ref{Fig_Phase_Diagrams}a, the LG coexistence exhibits a ``bottleneck'' in the vicinity of the LL critical temperature. This feature is characteristic of systems where polyamorphism is induced by interconversion, a result of the interaction between the thermodynamic path selected by interconversion and the critical line~\cite{Caupin_Thermodynamics_2019,Caupin_Minimal_2021,Longo_Interfacial_2023}.

The scaling theory predicts that, in the vicinity of the critical point, the coexistence curve is described by an asymptotic power law of the form~\cite{Ch8_Hassan_Mesothermo_2010}
\begin{equation}\label{Eq_OrderParamScaling}
    \varphi = \pm B_0|\Delta\hat{T}|^\beta
\end{equation}
where $\varphi$ is the order parameter for the phase transition, $\Delta\hat{T} = (T-T_\text{c})/T_\text{c}$ is the reduced distance to the critical temperature ($T_\text{c}$), $B_0$ is the critical amplitude, $\beta$ is the critical exponent, and the symbol ``$\pm$'' corresponds to each branch of the coexistence. 

\begin{figure}[t!]
    \centering
    \includegraphics[width=0.49\linewidth]{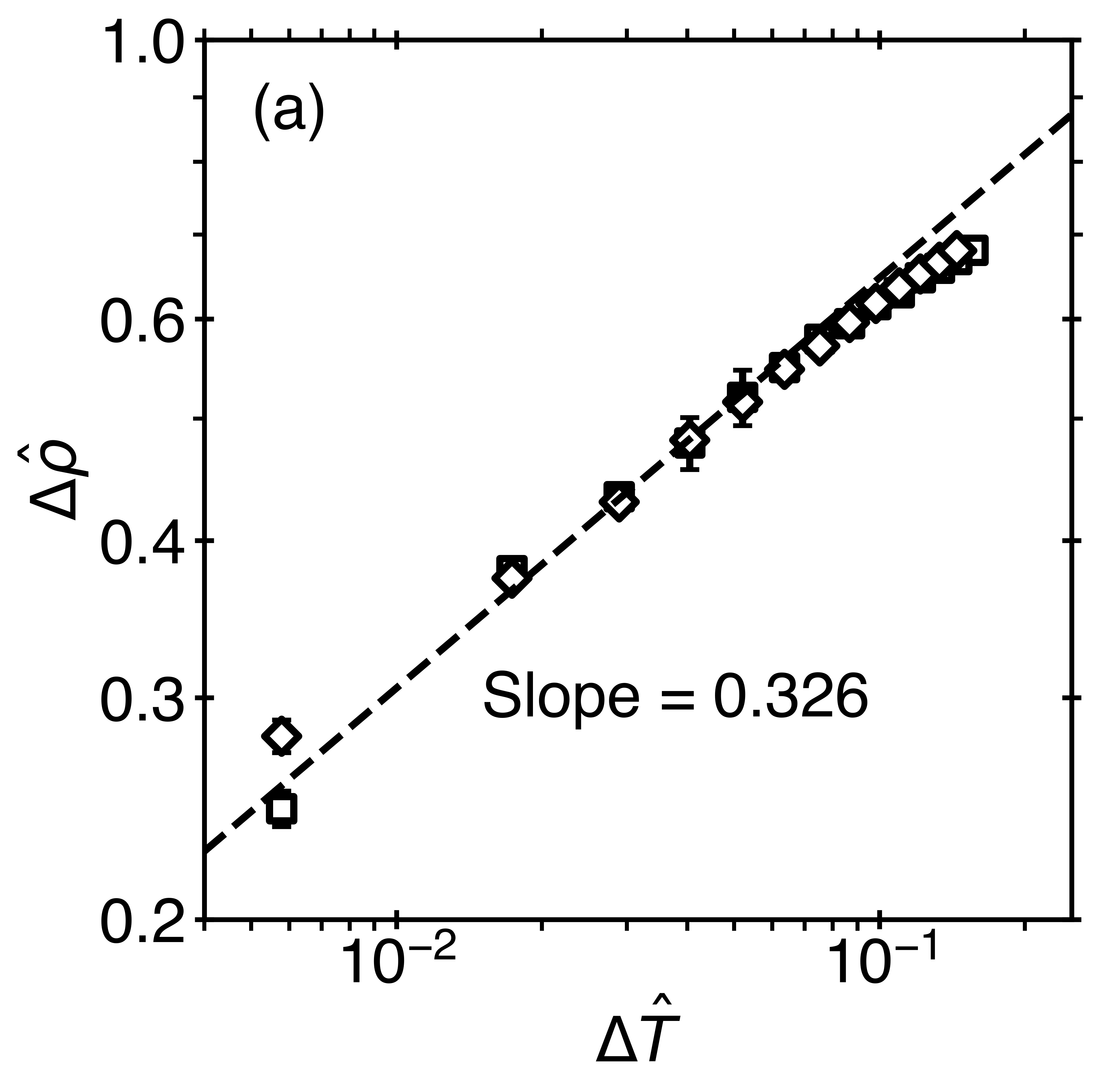}
    \includegraphics[width=0.49\linewidth]{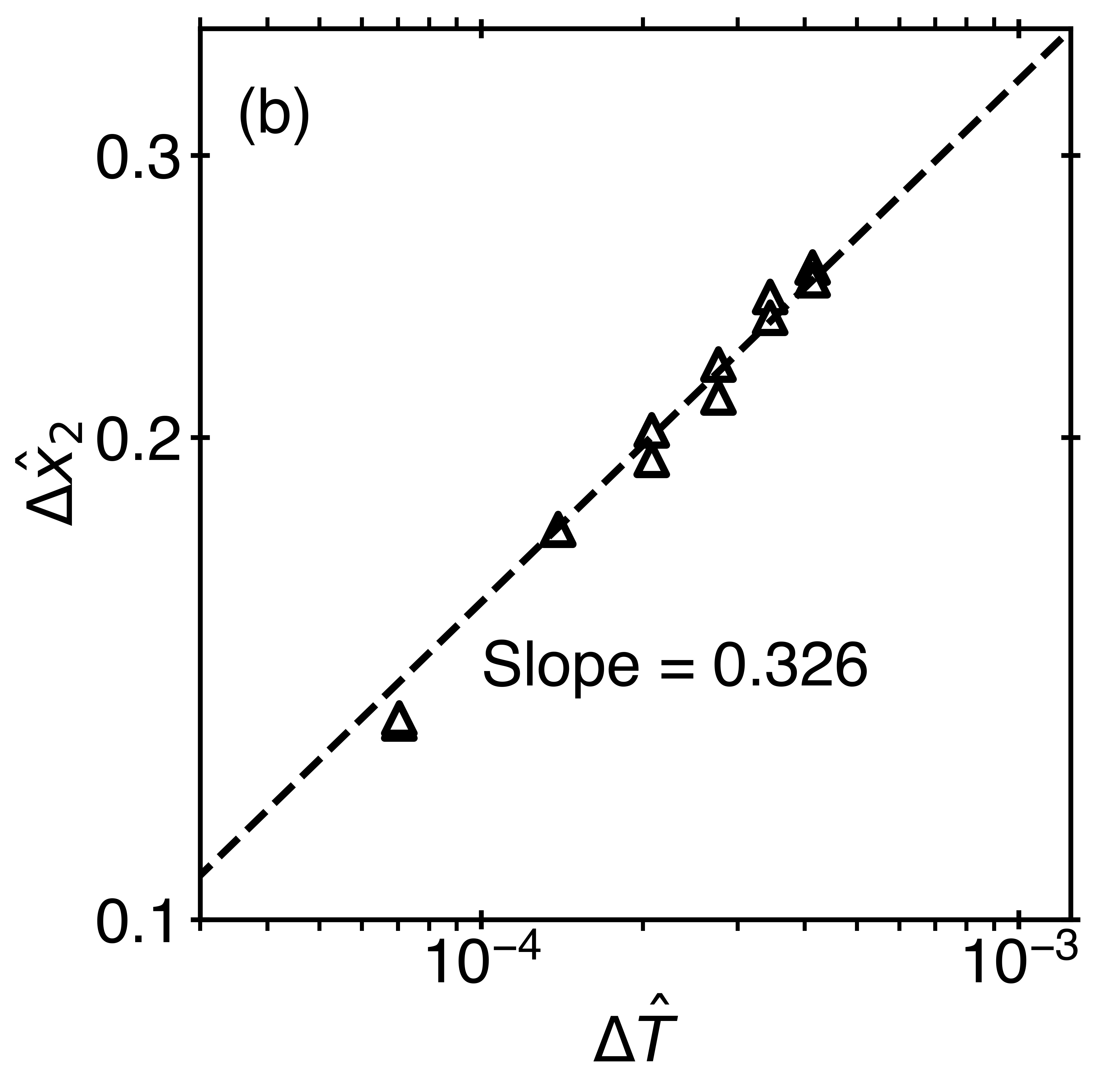}
    \caption{LG and LL phase coexistence in the vicinity of (a) the LGCP and (b) the LLCP. The density (a) and molecular fraction (b) for the system with $\omega = 1.4$ and $\ell = 256$ follow the predictions of the scaling theory (dashed line). In (a) the liquid and vapor branches are indicated with diamonds and squares, respectively. Note that in (a,b) the data point closest to the critical temperature is most strongly affected by the finite-size effect, causing the apparent deviation from the scaling power law.}
    \label{Fig_Red_Phase_Diagrams}
\end{figure}

Like similar lattice models~\cite{Shum_Phase_2021}, this system is expected to be in the three-dimensional Ising model universality class; the critical exponents for which are presented in Table~\ref{Table_Critical_Exponents}. For a binary mixture, in the vicinity of the LLCP, the order parameter is generally a nonlinear combination of the density and the molecular fraction~\cite{Wang_isomorphism_2008}. For an almost incompressible binary mixture, asymptotically close to the critical point, the density and the molecular fraction may be uncoupled and described by power laws given by
\begin{align}
    \Delta\hat{\rho} &= \pm B_{\rho,0}|\Delta\hat{T}|^\beta - B_{\rho,1}|\Delta\hat{T}|^{1-\alpha} \label{Eq_deltaRhoLL}\\
    \Delta\hat{x}_2 &= \pm B_{x,0}|\Delta\hat{T}|^\beta  \label{Eq_deltaXLL}
\end{align}
where $\Delta\hat{\rho} = (\rho - \rho_\text{c})/\rho_\text{c}$ and $\Delta\hat{x}_2 = (x_2 - x_{2,\text{c}})/x_{2,\text{c}}$ are the distances to the critical density, $\rho_\text{c}$, and the critical molecular fraction, $x_{2,\text{c}}$, respectively. From the fit of Eqs.~(\ref{Eq_deltaRhoLL}) and (\ref{Eq_deltaXLL}) to the MC simulation data, presented in Fig.~\ref{Fig_Phase_Diagrams}(c,d), the critical amplitudes of the density and the molecular fraction are obtained as, $B_{\rho,0} \approx -0.019$, $B_{\rho,1}\approx 3.02$, and $B_{x,0}\approx 3.192$. The ratio $B_{x,0}/B_{\rho,0} = \left(\partial \hat{P}/\partial \hat{T}|_{\hat{\rho}}\right)_\text{c}\approx -168$, where $\hat{P} = P/(\rho_\text{c}k_\text{B}T_\text{c})$, $\hat{T} = T/T_\text{c}$, and $\hat{\rho} = \rho/\rho_\text{c}$ are the reduced (by the corresponding critical parameters) pressure, temperature, and density, respectively, and the subscript ``$\text{c}$'' indicates that the derivative is to be evaluated at the critical point. 

\begin{figure}[t!]
    \centering
    \includegraphics[width=0.49\linewidth]{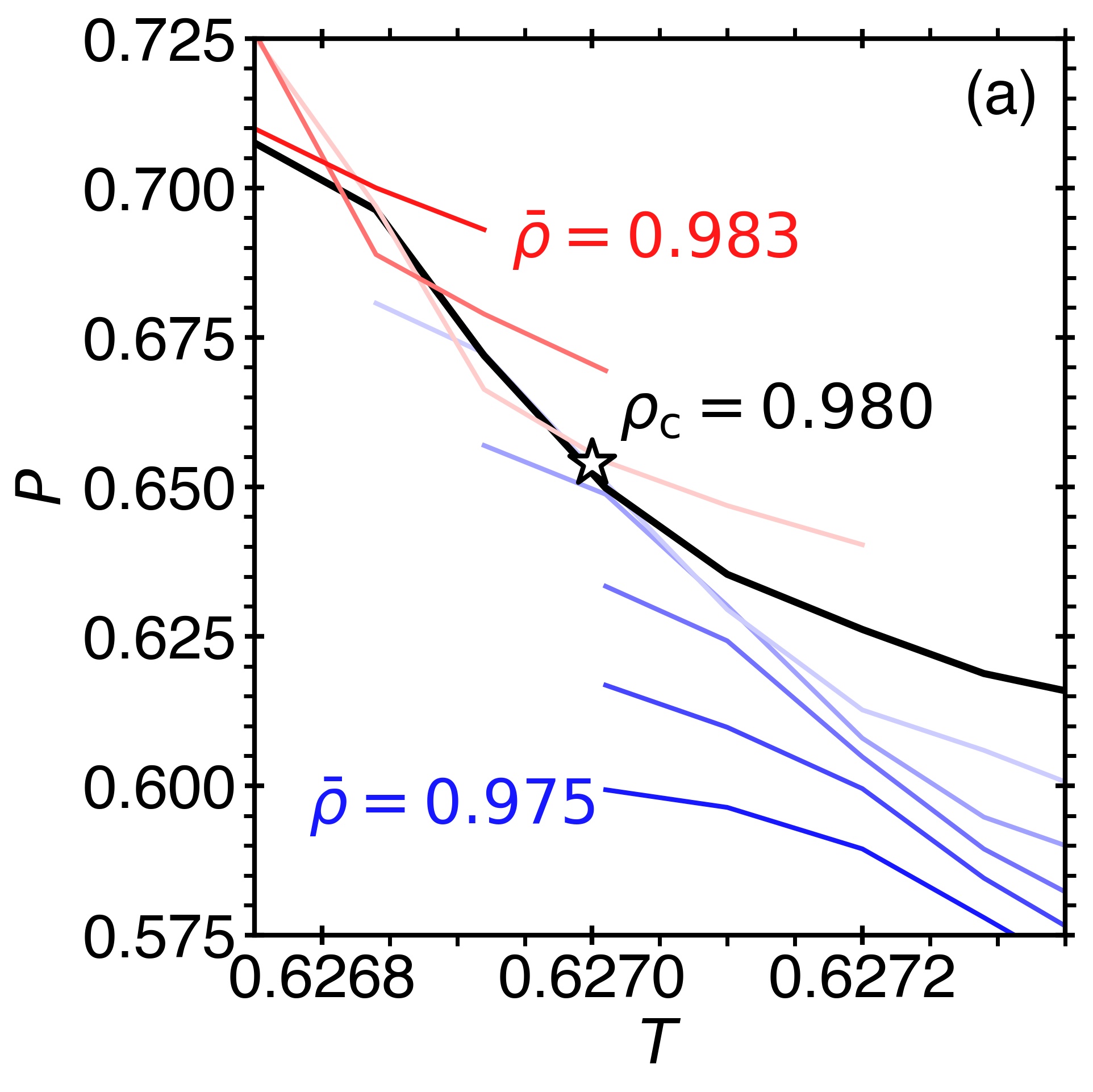}
    \includegraphics[width=0.49\linewidth]{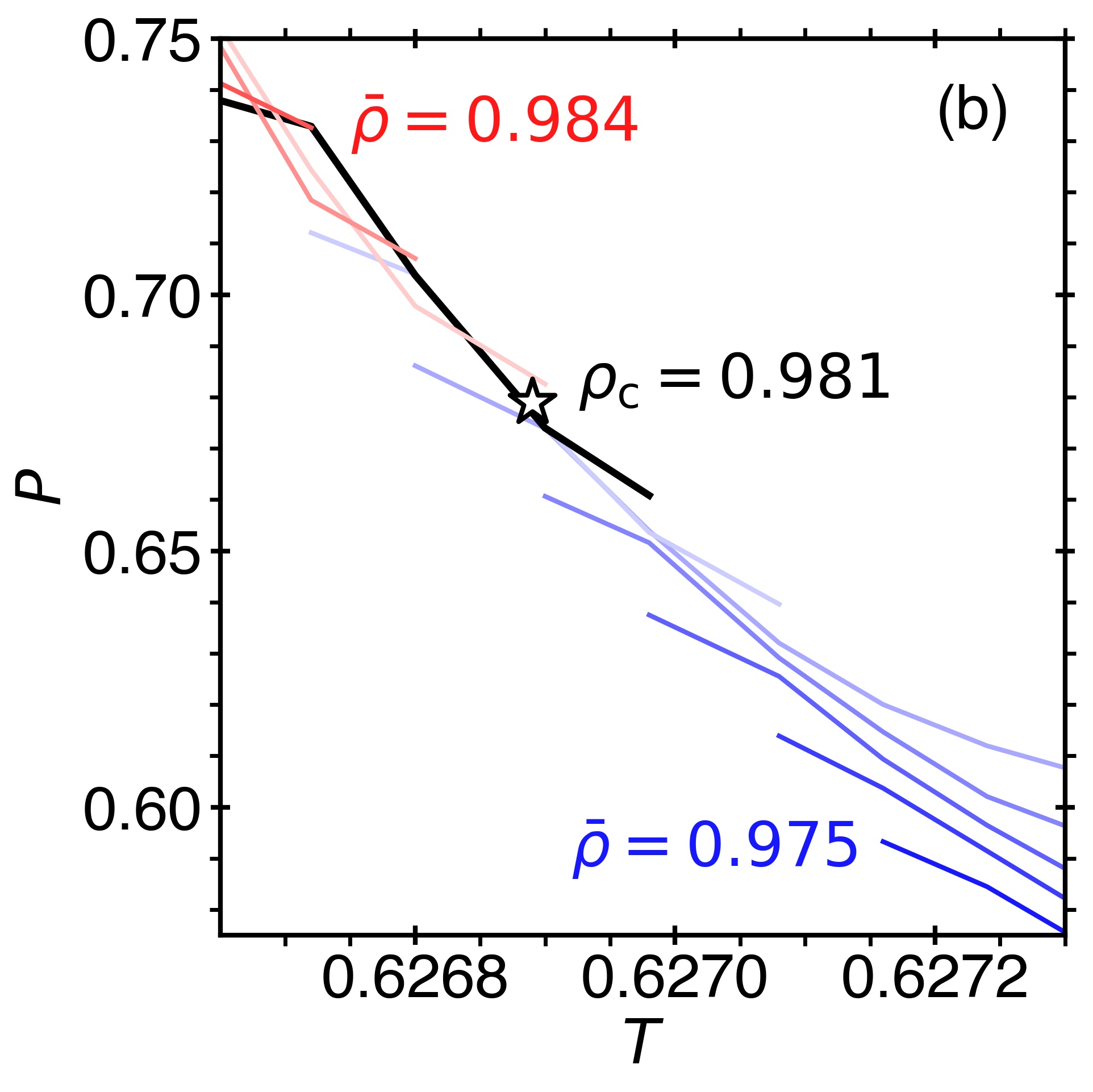}
    \caption{Pressure-temperature isochores near the LLCP for systems of sizes
    (a) $\ell=32$ and (b) $\ell=64$. Isochores are given in steps of the coexisting density of the two liquid phases $ \Delta\bar{\rho} = 0.001$ in which those with blue shades are below $\rho_\text{c}$, black is the critical isochore ($\rho_\text{c}$), and those with red shades are above $\rho_\text{c}$. The critical point (star) is approximately determined from the highest temperature and pressure of the intersecting isochores.}
    \label{Fig_Pressure}
\end{figure}

\begin{table}[b!]
\centering
\caption{Location of the LGCP and LLCP for systems of different sizes as predicted by the fit of the correlation length to Eq.~(\ref{Eq_Corr_Scaling}). In the fit, $\rho_\text{c}$ was selected and fixed based on the correlation length data with the largest value. The value for $T_\text{c}(\ell\to\infty)\equiv T_\text{c}^\infty$ was predicted from the extrapolation of the size-dependent values of $T_\text{c}$, based on the finite-size scaling theory~\cite{Fisher_finite_1972}, given by Eq.~(\ref{Eq_FiniteSize_Tc}) and shown in the inset of Fig.~\ref{Fig_LLcorr}. The value for $\rho_\text{c}(\ell\to\infty)$ was estimated by an empirical extrapolation to $\ell^{-1}$ of the size-dependent values of $\rho_\text{c}$.}
\label{Table_Critical_Points}
\begin{tabular}{cccc}
\toprule
Critical point & $\ell$ & $\rho_\text{c}$ & $T_\text{c}$ \\ \midrule
LGCP           & 256         & 0.5290           & 0.8605        \\
LLCP           & 64          & 0.9780           & 0.62683(1)      \\
  $-$             & 128         & 0.9790           & 0.62681(1)      \\
  $-$             & 256         & 0.9810           & 0.62680(1)    \\ 
  $-$           & $\infty$      & 0.9815           & 0.62679 \\ \bottomrule  
\end{tabular}
\end{table}

As indicated by Fig.~\ref{Fig_Phase_Diagrams}(c,d), while the LL coexistence is asymmetric in the density-temperature plane it is symmetric in the molecular fraction-temperature plane. This symmetry reflects that $x_2$ is the dominant contribution to the order parameter in the vicinity of the LLCP. Figure~\ref{Fig_Red_Phase_Diagrams} illustrates the LG and LL coexistence in the vicinity of the LGCP and LLCP, respectively, through the dominant contributions to the order parameter, the density and molecular fraction, given by Eqs.~(\ref{Eq_OrderParamScaling}) and (\ref{Eq_deltaXLL}). The slope of the curve in the vicinity of the critical point confirms the prediction of scaling theory, $\beta = 0.326$~\cite{Ch8_Hassan_Mesothermo_2010} for both critical points. The error bars depicted in Fig.~\ref{Fig_Red_Phase_Diagrams} were calculated by averaging over $N=5$ realizations. Note that the data point closest to the critical point is most affected by the finite-size of the system, producing the apparent deviation from the asymptotic power law. 

Pressure-temperature isochores in the vicinity of the LLCP, as depicted in Fig.~\ref{Fig_Pressure}, also yield an estimate for the location of the critical point. As mentioned in Section~\ref{Sec_WidomMethod}, the pressure calculations, based on the adjusted Widom-insertion method, are computationally expensive, and hence, they are only performed in the narrow region of the phase diagram close to the critical point for small system sizes: $\ell=32$ and $\ell=64$. One can see that both systems near the critical point exhibit a negative $\left(\partial \hat{P}/\partial \hat{T}|_{\hat{\rho}}\right)_\text{c}$ with an order of magnitude of approximately $-300$, which is near the value $-168$, predicted from the ratio of the amplitudes $B_{\rho,0}$ and $B_{x,0}$ for a larger size ($\ell = 256$). Also, from Fig.~\ref{Fig_Phase_Diagrams}d and Fig.~\ref{Fig_Pressure}, it is clear that the apparent critical temperature shifts towards lower values for increasing system sizes, which is consistent with the results obtained by the studies of the phase equilibrium. The location of the apparent (size-dependent) critical parameters for the systems investigated in this study are presented in Table~\ref{Table_Critical_Points}.

\subsection{Isothermal Compressibility and Correlation Length}

The structure factor was calculated using the method described in Ref.~\cite{Hansen_Liquid_2013} as
\begin{equation}
    \mathcal{S}_j({\bf q})=\frac{1}{n_j}\sum_{i=0}^n p_j({\bf q}) p_j({\bf -q})
\end{equation}
where $p_j({\bf q}) = \sum_{i=0}^n\delta_{js(i)}\exp({\bf q} \cdot {\bf r}_i)$ is the discrete Fourier transform of the positions of particles of type $j$ (in which $\delta_{js_i}$ is the Kronecker symbol), $\textbf{q}$ is a wave vector with components $q\equiv 2\pi\langle n_x,n_y,n_z\rangle/\ell$, in which $n_x$, $n_y$, and $n_z$ are integers between $-\ell/2$ and $\ell/2$ for each coordinate direction, $r_i$, respectively. The structure factor, $\mathcal{S}_j({\bf q})$, was averaged over radial bins of $\Delta q=2\pi/\ell$ and over $N=10$ realizations.

\begin{figure}[b!]
    \centering
    \includegraphics[width=0.49\linewidth]{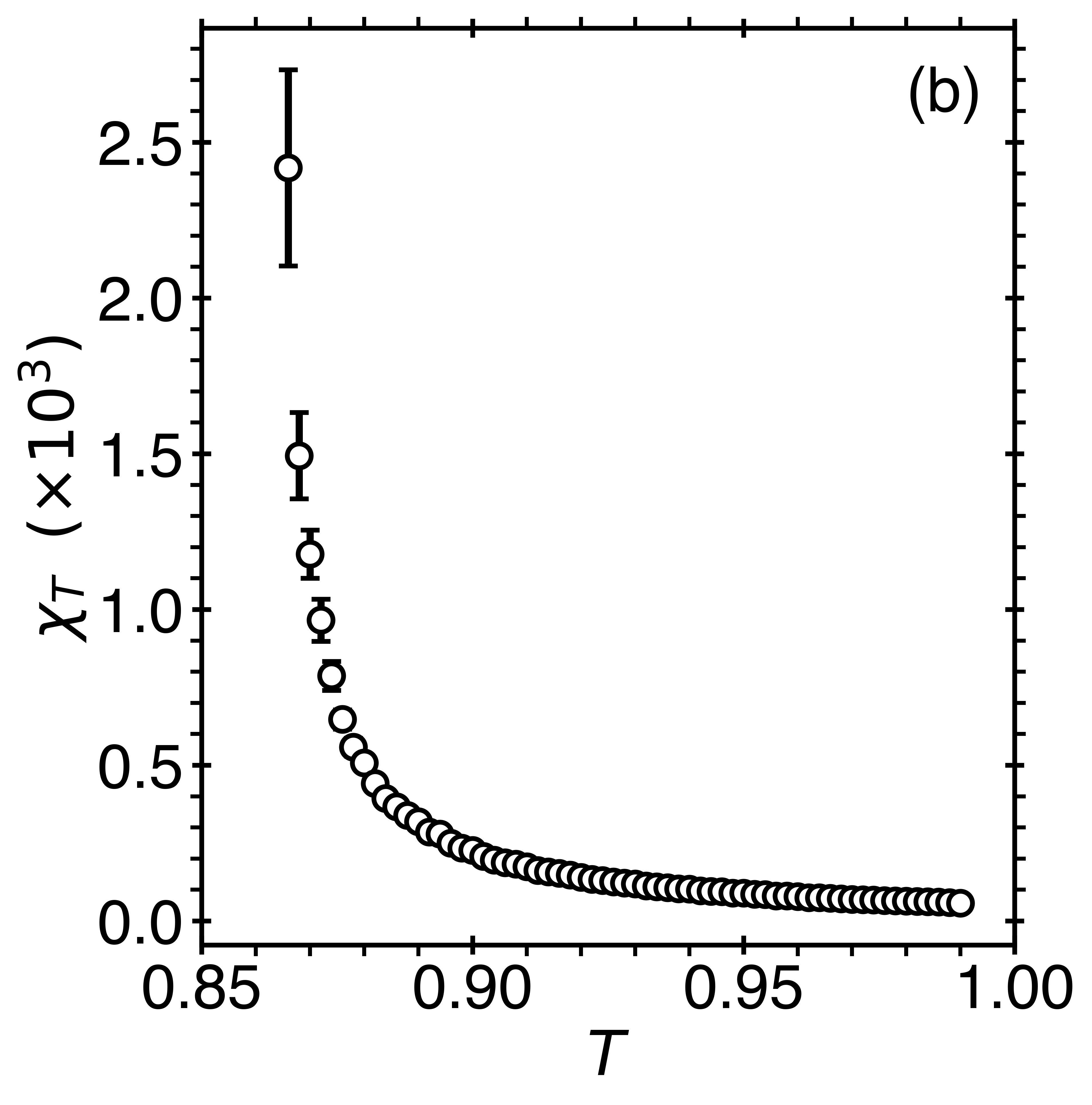}
    \includegraphics[width=0.49\linewidth]{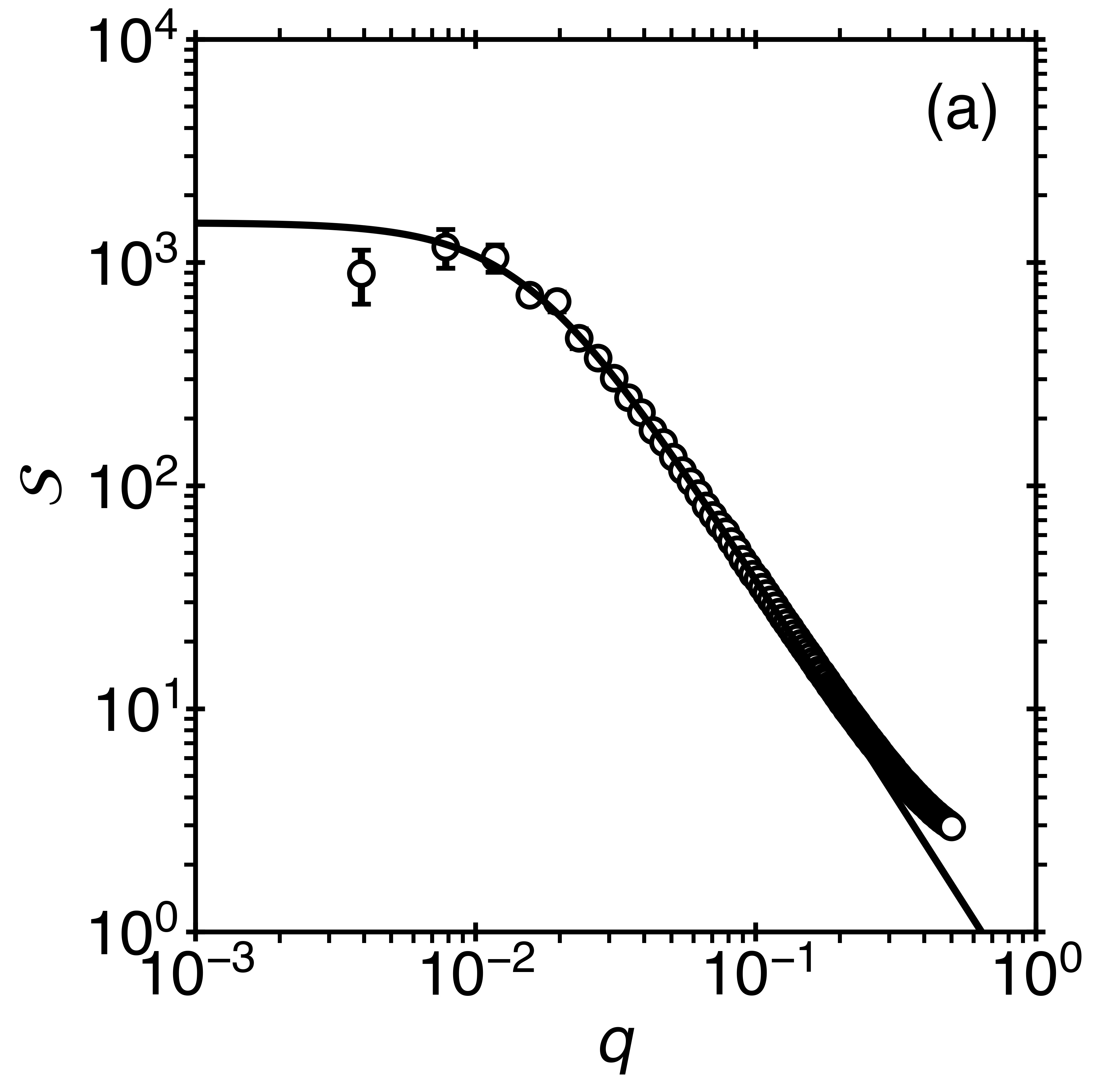}
    \caption{Structure factor (a) and inverse structure factor (b) in the vicinity of the LGCP for a system with $T=0.868$, $\bar{\rho} = 0.53$, and $\ell = 256$. The open circles correspond to MC simulations averaged over $N=10$ realizations, while the curve corresponds to the theoretical prediction of Eq.~(\ref{Eq_Sfactor}).}
    \label{Fig_Struct_Fact}
\end{figure}

\begin{figure}[t!]
    \centering
    \includegraphics[width=0.49\linewidth]{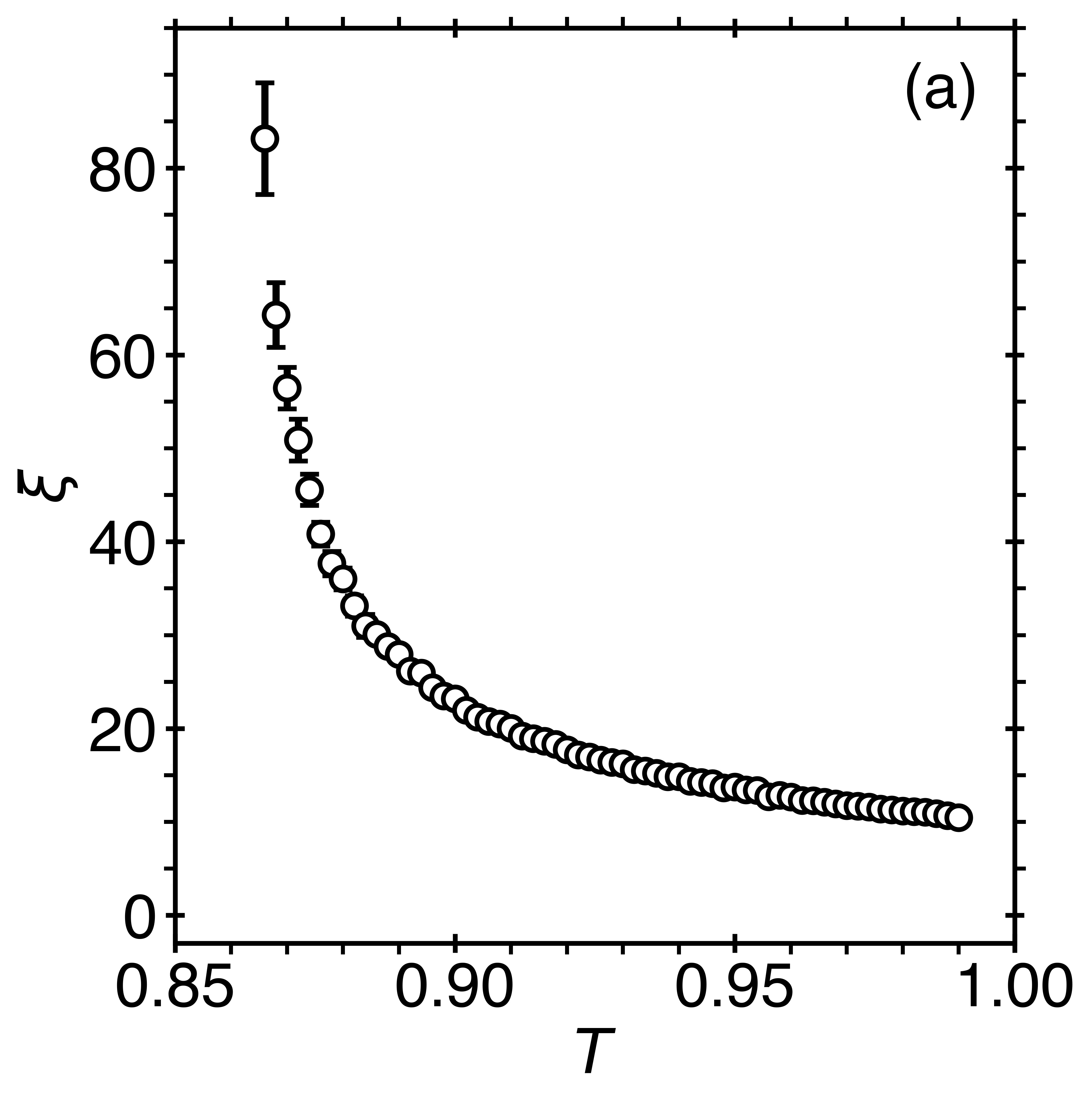}
    \includegraphics[width=0.49\linewidth]{SusceptLG.jpg}
    \includegraphics[width=0.49\linewidth]{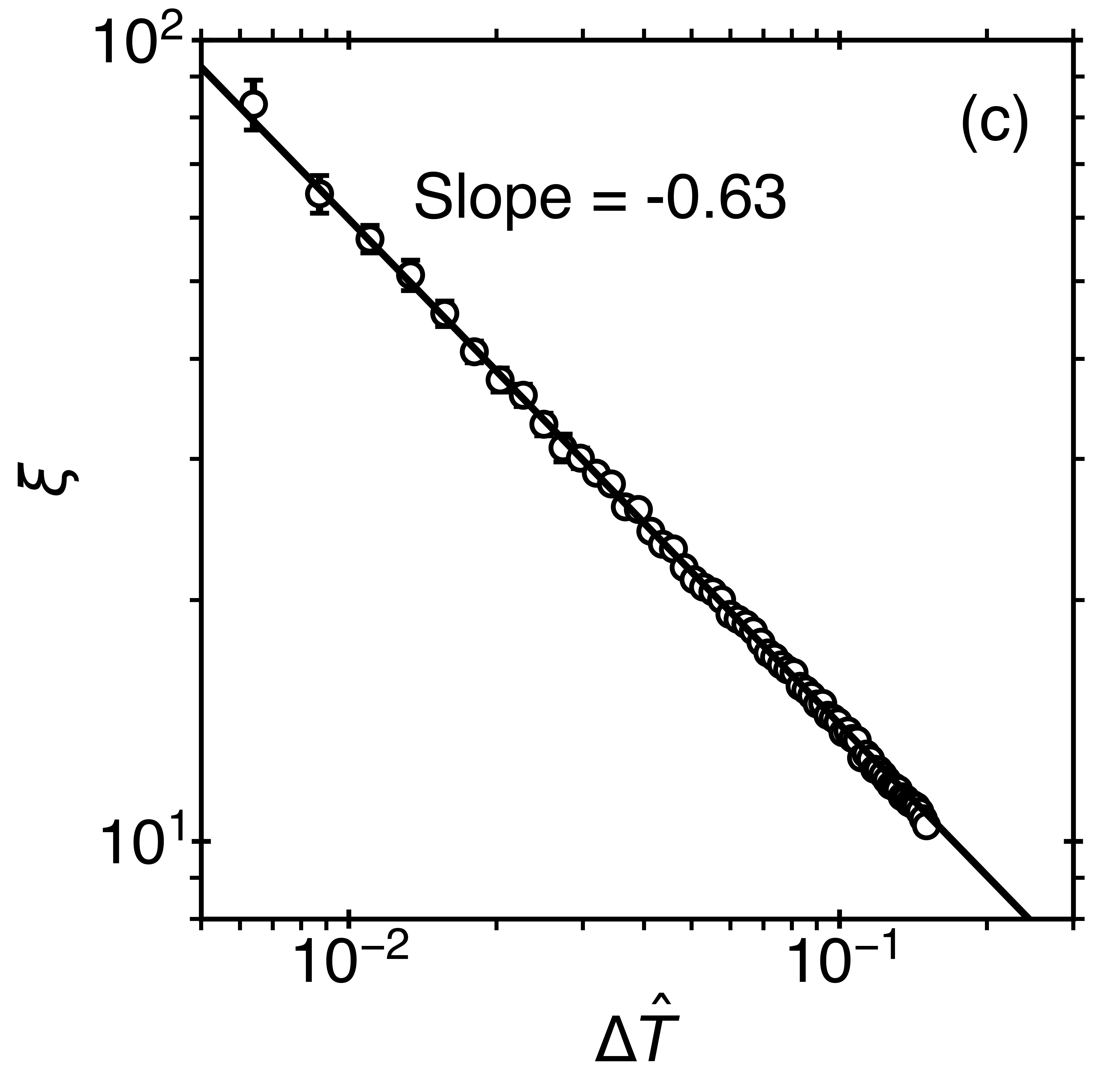}
    \includegraphics[width=0.49\linewidth]{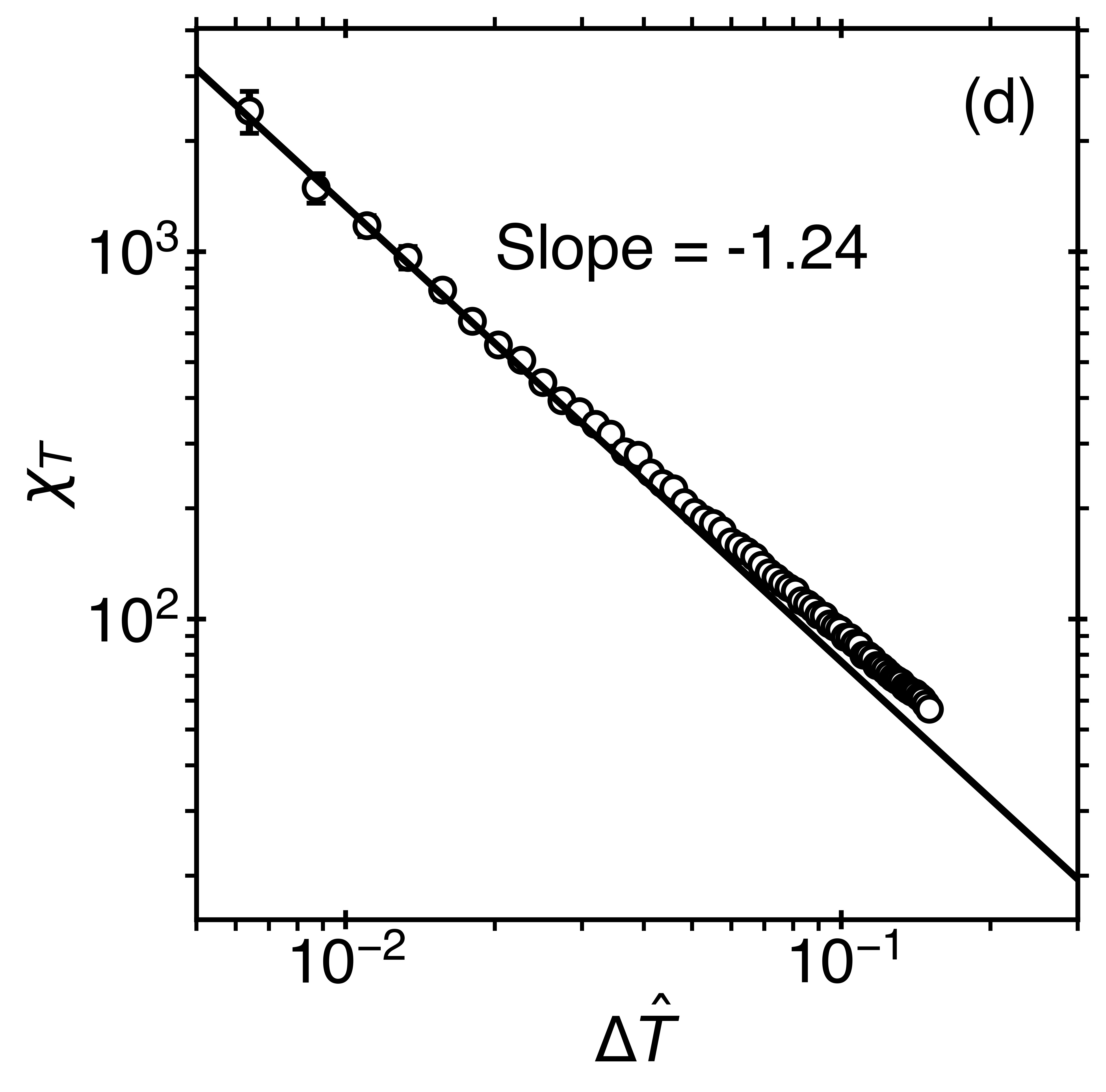}
    \caption{Correlation length (a) and isothermal compressibility (b) extracted from the LG structure factor data for the same system considered in Fig.~\ref{Fig_Struct_Fact} with $T=0.868$, $\bar{\rho} = 0.53$, and $\ell = 256$. In the vicinity of the LGCP, the correlation length (c) and the isothermal compressibility (d) follow the theoretical prediction of the finite-size scaling theory. One may also notice deviations of the susceptibility from the asymptotic power law, given by Eq.~(\ref{Eq_Sus_Scaling}), away from the critical point, which are associated with the crossover to the meanfield behavior~\cite{Kim_Crossover_2003}.}
    \label{Fig_Corr_Len}
\end{figure}

Figure~\ref{Fig_Struct_Fact} depicts the simulation results, performed in a cubic box of length $\ell$, of the structure factor as a function of the wavenumber, $q = |\textbf{q}|$, in the vicinity of the LGCP. The structure factor is related to the size-dependent isothermal compressibility (susceptibility), $\chi_T = \chi_T(\xi,\ell)$, multiplied by the correlation function, $G(q\xi)$ as~\cite{Ch8_Kostko_Crossover_2005}, 
\begin{equation}\label{Eq_Sfactor}
    \mathcal{S}(q, \xi,\ell) = \chi_T(\xi,\ell)G(q\xi)
\end{equation}
where $\xi$ is the correlation length and $\ell$ is the size of the system. Typically, the Ornstein-Zernike approximation for the correlation function, $G(q\xi) = 1/\left[1+(q\xi)^2\right]$ is sufficient to describe the structure factor. However, in the vicinity of the critical point, when $q\xi \gg 1$, the Ornstein-Zernike becomes insufficient as $G(q\xi)\sim (q\xi)^{2-\eta}$. A simplified version of the Fisher-Burford approximation~\cite{Fisher_scattering_1967} that satisfies this scaling requirement can be expressed as
\begin{equation}\label{Eq_FisherBurford_Approx}
    G(q\xi) \approx \left[1 + (q\xi)^2\right]^{\eta/2 - 1}
\end{equation}
The fit of the structure factor to Eq.~(\ref{Eq_Sfactor}), with use of the simplified Fisher-Burford approximation, is illustrated by the black curves on Fig.~\ref{Fig_Struct_Fact}(a,b). 

From the structure factor, the size-dependent susceptibility and the correlation length of order parameter fluctuations may be determined. Scaling theory predicts that the size-independent correlation length and susceptibility (for an infinite-sized system), asymptotically close to the critical point, follow power laws of the form
\begin{align}
    \xi &= \xi_0|\Delta\hat{T}|^{-\nu} \label{Eq_Corr_Scaling}\\
    \chi_T &= \Gamma_0|\Delta\hat{T}|^{-\gamma}\label{Eq_Sus_Scaling}
\end{align}
where $\xi_0$ and $\Gamma_0$ are the critical amplitudes of the correlation length and susceptibility, respectively, while $\nu$ and $\gamma$ are the corresponding critical exponents (see Table~\ref{Table_Critical_Exponents}). Note that these two properties are related as $\chi_T(\xi) \sim \xi^{\gamma/\nu}$ since $\gamma = \nu(2-\eta)$.

However, as the susceptibility is strongly dependent on the size of the system, finite-size effects cause deviations as $\chi_T=\chi_T(\xi)$ when $\xi/\ell\ll 1$, while $\chi_T=\chi_T(\ell)$ when $\xi/\ell\gg 1$. A form of the susceptibility that satisfies these limits is given by
\begin{equation}\label{Eq_FiniteSize_Suscept}
    \chi_T(\xi,\ell) = \chi_T(\xi)\Upsilon(\xi,\ell)
\end{equation}
where the finite-size scaling function can be approximated as
\begin{equation}
    \Upsilon(\xi,\ell) = \left[1 + \left(\frac{2\xi}{\ell}\right)^2\right]^{-\gamma/(2\nu)} 
\end{equation}
In the thermodynamic limit $\xi/\ell\to 0$, such that $\Upsilon\to 1$ and the susceptibility is given by Eq.~(\ref{Eq_Sus_Scaling}). In the finite-size limit $\xi/\ell\to\infty$, such that $\Upsilon\to (2\xi/\ell)^{-\gamma/\nu}$, hence the susceptibility is given by $\chi_T =(\ell/ 2)^{\gamma/\nu}$. The critical amplitudes for the systems studied are given in Table~\ref{Table_Critical_Amplitudes}. In conceptual agreement with the finite-size scaling theory, the amplitudes, within their error bars, are independent of the system size.  

With use of Eq.~(\ref{Eq_Sfactor}) for the structure factor, the  correlation length and the size-dependent susceptibility may be extracted from the structure factor data. Results of this process in the vicinity of the LGCP are presented in Fig.~\ref{Fig_Corr_Len}. The susceptibility and correlation length follow asymptotic scaling laws typical of the three-dimensional Ising-like universality class as illustrated in Fig.~\ref{Fig_Corr_Len}(c,d). Note that, for the large system of size of $\ell=256$, since the simulation data are relatively far from the LGCP, the effect of finite-size is not observed in Fig.~\ref{Fig_Corr_Len}d. Instead, one may notice that the susceptibility deviates from the asymptotic scaling power law away from the critical point, which occurs since this region ($\Delta\hat{T}\gtrsim 0.1$) is associated with the crossover to the meanfield behavior. This observation is consistent with other MC simulations of the three-dimensional Ising model with a short range of interactions ($z=6$)~\cite{Kim_Crossover_2003}. 

\begin{figure}[t!]
    \centering
    \includegraphics[width=0.49\linewidth]{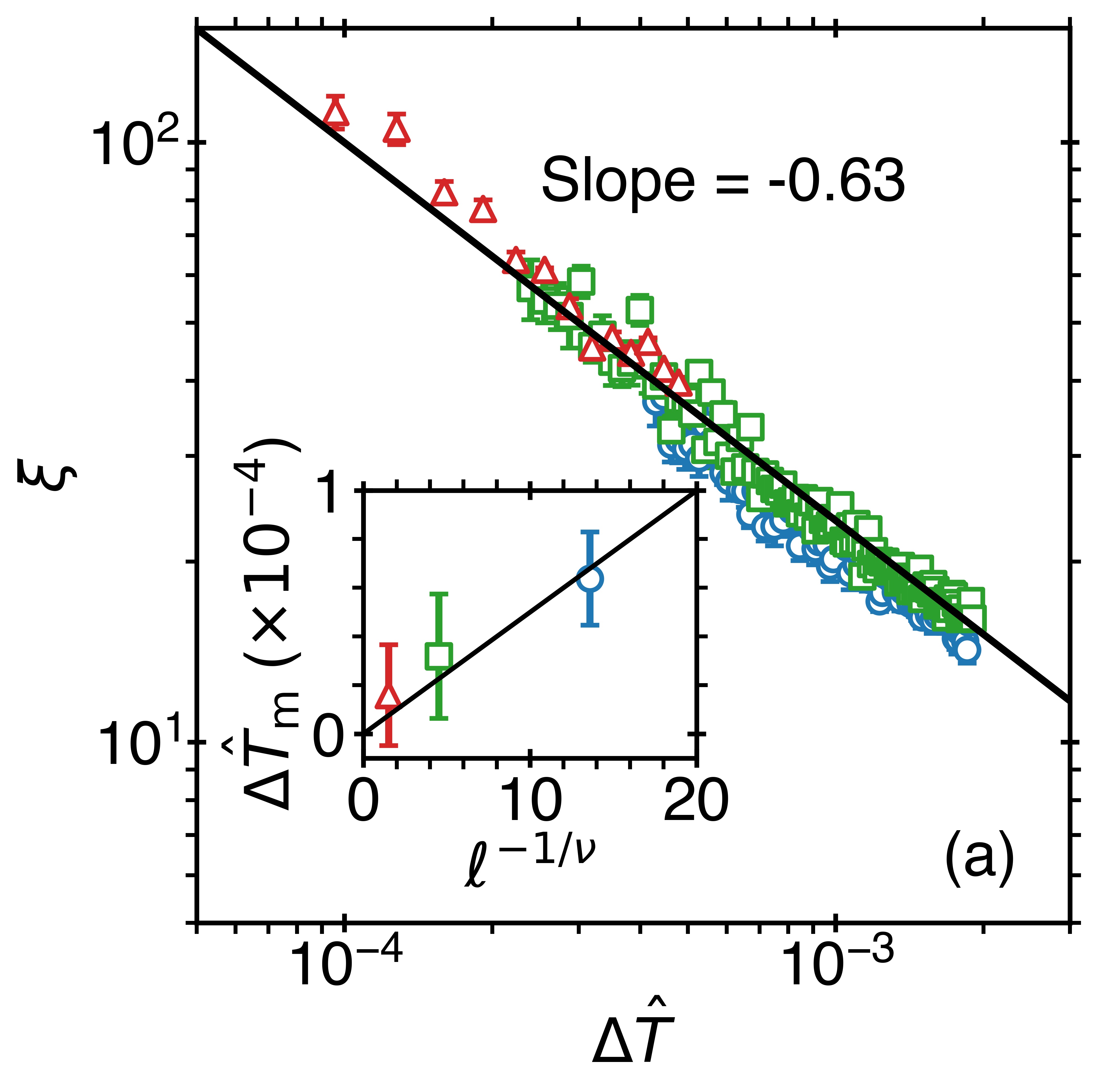}
    \includegraphics[width=0.49\linewidth]{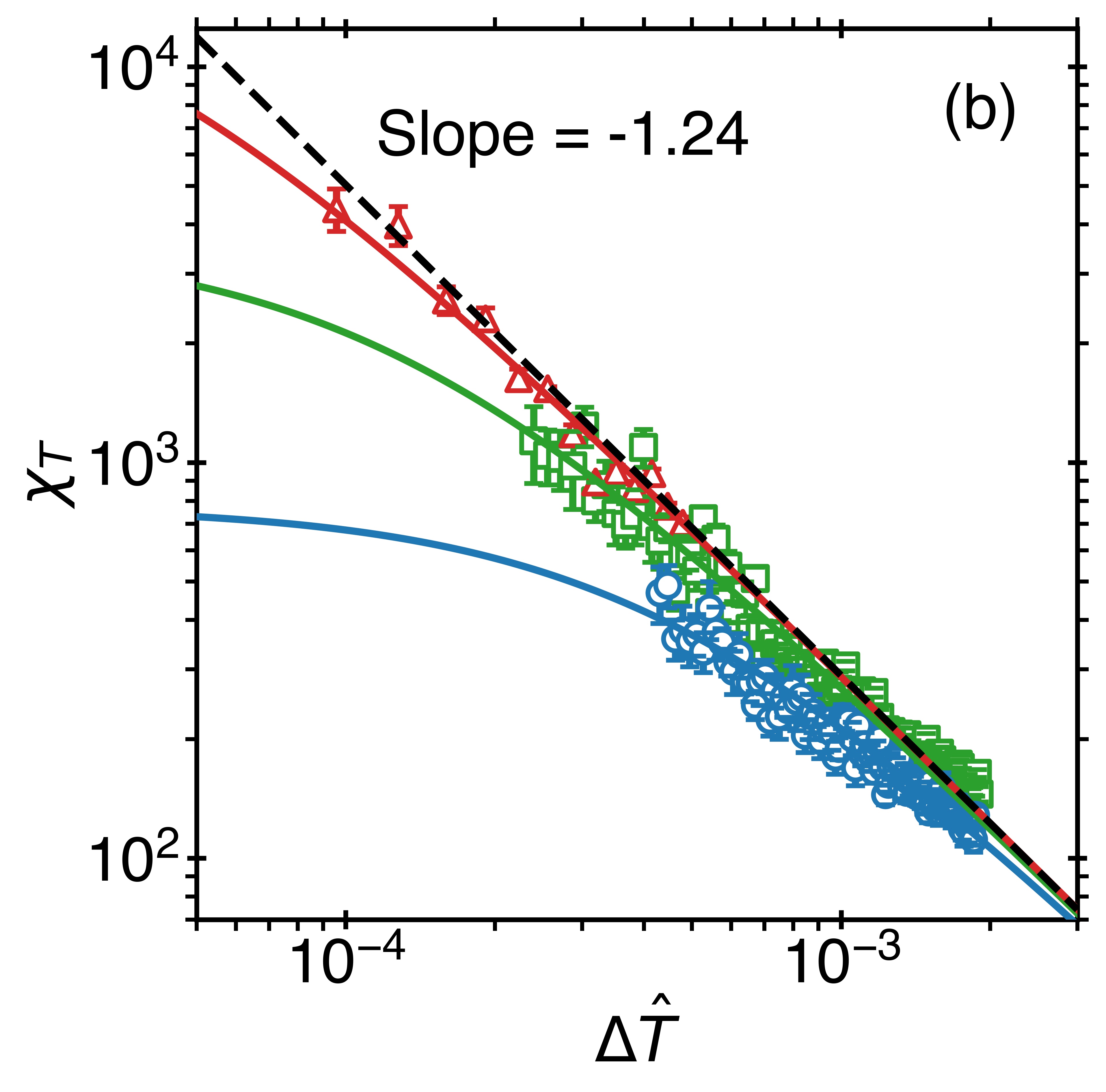}
    \caption{Correlation length (a) and susceptibility (b) in the vicinity of the LLCP. The effects of finite size are demonstrated by the different simulations at $\ell = 64$ (blue circles), $\ell = 128$ (green squares), and $\ell = 256$ (red triangles). (a) The solid line is the prediction of the correlation length from scaling theory, Eq.~(\ref{Eq_Corr_Scaling}). The inset of illustrates that the change in the critical temperature due to the finite-size effect is in agreement with the prediction of the scaling theory, Eq.~(\ref{Eq_FiniteSize_Tc}). (b) The solid curves depict the predictions of the susceptibilities from the finite-size scaling theory, Eq.~(\ref{Eq_FiniteSize_Suscept}). Note that away from the critical point, the effects of finite size observed in the susceptibility diminish as all curves converge.}
    \label{Fig_LLcorr}
\end{figure}

Results of extracting the correlation length and susceptibility from the structure factor in the vicinity of the LLCP for systems of three different sizes are presented in Fig.~\ref{Fig_LLcorr}. The parameters from the fit of Eq.~(\ref{Eq_FiniteSize_Suscept}) to the susceptibility data are provided in Table~\ref{Table_Critical_Amplitudes}. As illustrated on Fig~\ref{Fig_LLcorr}b, the susceptibility for the smallest size considered, $\ell = 64$, is the most strongly affected by the finite-size effect, significantly deviating from the predictions of Eq.~(\ref{Eq_Sus_Scaling}), while the susceptibility for the largest size considered, $\ell = 256$, is only slightly affected. Note that the simulation data presented in Fig.~\ref{Fig_LLcorr} are much closer to the LLCP than the simulation data presented in Fig.~\ref{Fig_Corr_Len}, such that the finite-size effect may be readily observed for all system sizes. 

The finite-size scaling theory predicts that the apparent critical temperature, $T_\text{c}(\ell)$, given in Table~\ref{Table_Critical_Points}, obtained from Eq.~(\ref{Eq_FiniteSize_Suscept}), scales with the system size as~\cite{Fisher_finite_1972} \begin{equation}\label{Eq_FiniteSize_Tc}
    \Delta\hat{T}_\text{m} = \frac{T_\text{c}(\ell)}{T_\text{c}^\infty} - 1 \sim \ell^{-1/\nu}
\end{equation}
where $T_\text{c}^\infty$ is the critical temperature of a system with infinite size. The inset on Fig.~\ref{Fig_LLcorr}a confirms this prediction. Additionally, from the extrapolation of the apparent critical temperatures, it is predicted that $T_\text{c}^\infty = 0.62679$.

Note that there is a small discrepancy between the prediction of the locations for the LGCP and LLCP obtained from the phase coexistence, presented in Figs.~\ref{Fig_Phase_Diagrams} and \ref{Fig_Pressure}, with those obtained from the structure factor, presented in Table~\ref{Table_Critical_Points}. For a system of size $\ell = 256$, the critical temperature obtained from coexistence is $T_\text{c}^\text{LG} = 0.8605$ and $T_\text{c}^\text{LL} = 0.62677(5)$, while from the structure factor $T_\text{c}^\text{LG} = 0.8585$ and $T_\text{c}^\text{LL} = 0.62680(2)$. This effect requires further investigation. However, one may attribute this discrepancy (on the order of $10^{-4}$) to the different geometries used for simulations of phase coexistence (rectangular box) and structure factor (cubic box).We believe the prediction of the critical point from the correlation length to be the most accurate.

\begin{table}[t!]
\centering
\caption{Predictions from the finite-size scaling theory, given by Eq.~(\ref{Eq_FiniteSize_Suscept}), for the critical amplitudes $\xi_0$ and $\Gamma_0$ of the correlation length and susceptibility, respectively. These predictions were obtained from the fits to the susceptibility data and are presented in Figs.~\ref{Fig_Corr_Len} and \ref{Fig_LLcorr}.}
\begin{tabular}{cccc}
\toprule
Critical point & System size & $\xi_0$ & $\Gamma_0$ \\ \midrule
LGCP           & 256         & 3.284(2)   & 4.407(2)        \\
LLCP           & 64          & 0.245(7)   & 0.055(5)           \\
$-$               & 128         & 0.228(5)   & 0.055(3)          \\
$-$               & 256         & 0.190(3)   & 0.055(2)         \\ \bottomrule  
\end{tabular}
\label{Table_Critical_Amplitudes}
\end{table}

\subsection{Surface Tension}

In this subsection, the liquid-gas interfacial tension, $\sigma_\text{LG}$, is obtained from two methods. First, the surface tension was calculated via Binder's thermodynamic integration of the Helmholtz free energy~\cite{Binder_Calculation_1982}. An additional constraint, that the interfacial tension must be zero at the critical point, was incorporated as a constant shift in the surface tension calculated via this method. Second, the interfacial tension was obtained from calculations of the Laplace pressure. The Laplace pressure, $P_\mathcal{L}$, is the pressure difference between the inner and outer phases of a system consisting of a vapor bubble in equilibrium with a surrounding liquid phase. It is related to the surface tension as $P_\mathcal{L} = -2\sigma_\text{LG}/R$, in which $R$ is the radius of the bubble, calculated as 
\begin{equation}
    R=\ell\left(\frac{3}{4\pi}\frac{\rho_\text{L}-\bar{\rho}}{\rho_\text{L}-\rho_\text{G}}\right)^{1/3}
\end{equation}
where $\rho_\text{L}$ and $\rho_\text{G}$ are the equilibrium values of the densities of the liquid (outside the bubble) and the vapor (inside the bubble), respectively, while $\bar{\rho}=(n_1+n_2)/n$ is the fixed average density of the system. The Laplace pressure can be obtained from the difference between the liquid pressure for a system with a flat interface, $P_0$, and the liquid pressure for a system with a bubble, $P_\mathrm{L}$. A good approximation for the expression of $P_\mathcal{L}$ is given by the Poynting correction in the incompressible limit~\cite{Debenedetti_metastable_1996,Arvengas2011}: $P_\mathcal{L} = (P_0-P_{\text{L}})(1-\rho_{\rm G}/\rho_{\rm L})$. Finally, the interfacial tension is written as
\begin{equation}
    \sigma_{\rm LG}= \frac{2}{R}(P_0-P_{\text{L}})\left(1-\frac{\rho_{\rm G}}{\rho_{\rm L}}\right)
\end{equation}

\begin{figure}[b!]
    \centering
    \includegraphics[width=0.49\linewidth]{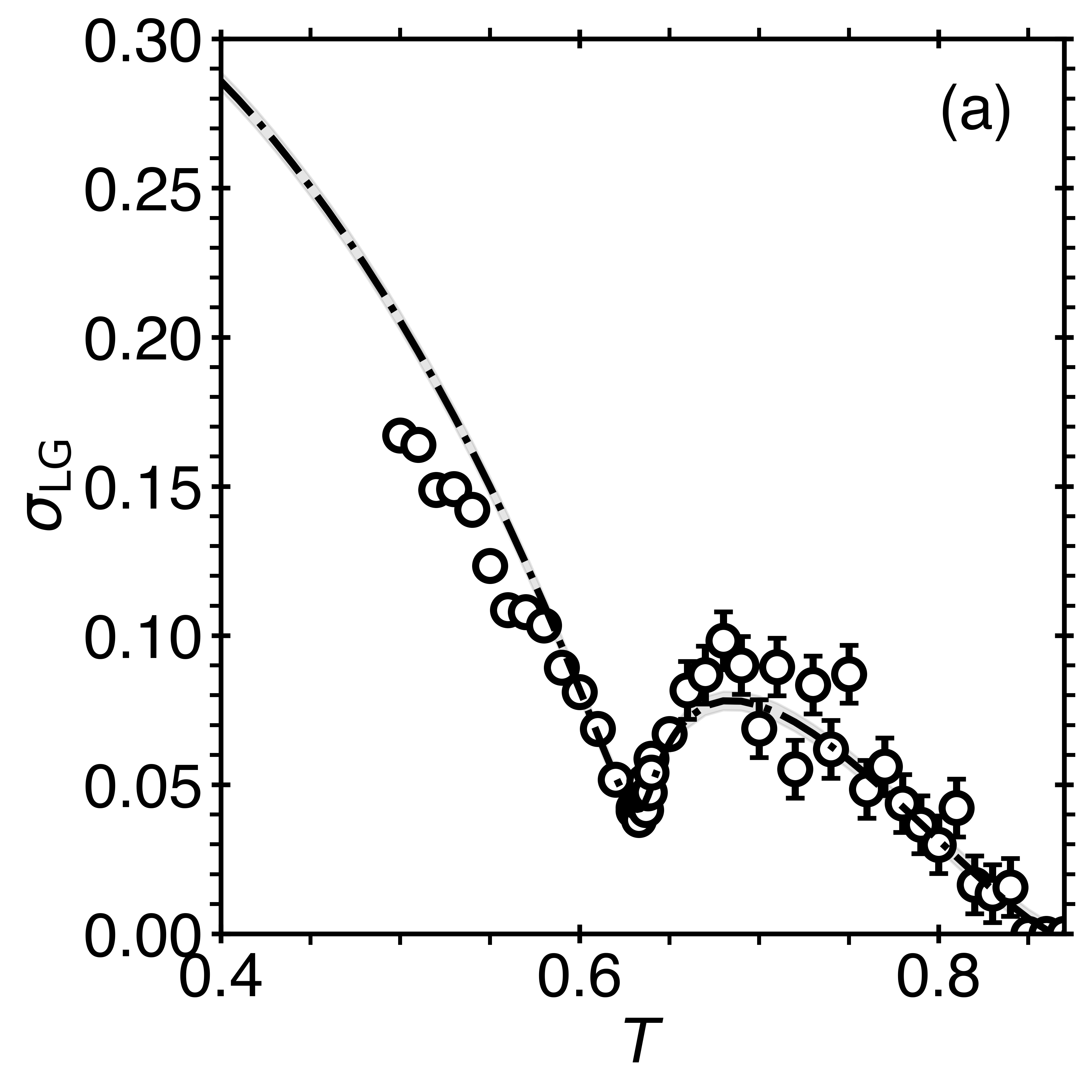}
    \includegraphics[width=0.49\linewidth]{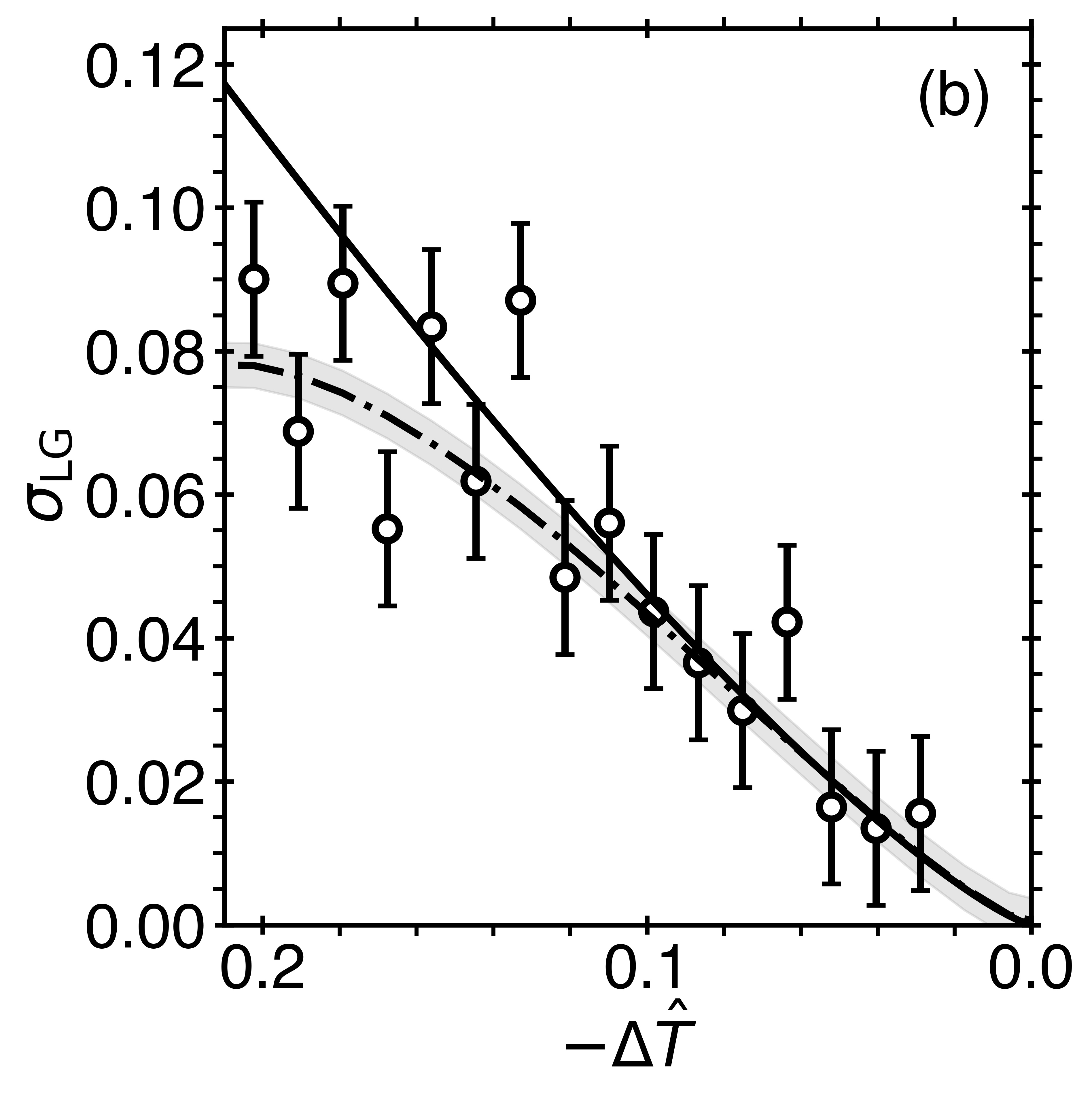}
    \caption{(a) Surface tension obtained from the methods of the Laplace pressure (open circles) and Binder's thermodynamic integration (dot-dashed curve). Both methods confirm the non-monotonic surface tension predicted by the meanfield theory (see details in Ref.~\cite{Longo_Interfacial_2023}). (b) In the vicinity of the liquid-gas critical point, both methods follow the predictions of the scaling theory (exponent $\mu=1.26$, solid line) to the limit of the estimated error.}
    \label{Fig_SigmaLG}
\end{figure}

Figure~\ref{Fig_SigmaLG}a displays the results for the liquid-gas surface tension, $\sigma_\text{LG}$, obtained from these two methods, demonstrating a reasonably good agreement. They both show an anomalous behavior: upon warming, the surface tension reaches a minimum, followed by a maximum, before vanishing at the LGCP. This surprising non-monotonic variation was already observed in our previous meanfield approach~\cite{Longo_Interfacial_2023} for sufficiently low values of $\omega_{12}$, and is confirmed by the present MC simulations. The extrema in surface tension occurs at temperatures close to the density extrema along LG coexistence, see Fig.~\ref{Fig_Phase_Diagrams}a.

To our knowledge, there is yet no other report of a surface tension minimum for a water-like system. A maximum in the surface tension was previously reported in a van der Waals-Cahn-Hilliard treatment of a water-like fluid, but a different assumption about the influence parameter involved in the treatment lead, instead, to a monotonic temperature dependence with an inflection point. Surface tension measurements in the deeply supercooled liquid suggest that water may possess such an inflection point~\cite{Vins_Surface_2020}. This property has received much attention throughout the years, with a number of predictions from molecular dynamics simulations with various water-like potentials~\cite{Lu_second_2006,Rogers_evidence_2016,Wang_second_2019,Malek_surface_2019,Gorfer_high_2023,Hrahsheh_second_2024}. The low-temperature inflection point in the surface tension also occurs in the blinking-checkers model for large enough $\omega_{12}$~\cite{Longo_Interfacial_2023} in the meanfield treatment; in that case, the fluid exhibits water-like anomalies but no liquid-liquid transition. It would be interesting, in a future work, to study this case with the exact calculation.

Scaling theory predicts that in the vicinity of the critical point, the surface tension is described by an asymptotic power law of the form~\cite{Widom_surface_1965}
\begin{equation}
    \sigma = \sigma_0|\Delta\hat{T}|^\mu
\end{equation}
where $\sigma_0$ is the critical amplitude of the surface tension, while $\mu$ is the critical exponent, being $\mu = 1.26$ for three-dimensional Ising-like systems (see Table~\ref{Table_Critical_Exponents}). Figure~\ref{Fig_SigmaLG}b shows $\sigma_\text{LG}$ in the vicinity of the LGCP, but due to the computational expense of the calculations, it is difficult to ascertain which method is more accurate. The two methods are in agreement with the scaling theory, to the limit of the estimated error, being characterized by a critical amplitude of $\sigma_0 \simeq 0.84$, indicated by the black curve in Fig.~\ref{Fig_SigmaLG}b.

\section{Conclusions}
In this study, Monte Carlo simulations of the blinking-checkers model~\cite{Caupin_Minimal_2021}
for polyamorphic fluids are performed, demonstrating the fluid phase behavior and confirming the predictions of the meanfield theory~\cite{Longo_Interfacial_2023}. In addition, Monte Carlo simulations provide access to information on the correlation length, which is not available from the meanfield prediction. Calculations of the coexistence, isothermal compressibility, correlation length, and surface tension from the simulation data demonstrate that both the liquid-gas critical point and the liquid-liquid critical point exhibit three-dimensional Ising-like critical behavior. The locations of these critical points are found to be different from their meanfield counterparts for this model with the same set of interaction parameters~\cite{Longo_Interfacial_2023}, as the behavior of the thermodynamic properties are significantly affected by critical fluctuations and finite-size effects.

\section*{Acknowledgements}
The authors would like to thank Nikolay Shumovskyi for his assistance with the initial simulations. F.C. and M.A. appreciate the beneficial collaborations with the scientists at Universidad Complutense de Madrid. In particular, F.C. thanks Jos\'{e} Luis F. Abascal for a fruitful collaboration and for his hospitality during visits in Madrid.

\section*{Disclosure statement}

The authors declare no conflicts of interest.

\section*{Funding}
 The research of S.V.B. was supported by NSF Award No. 1856496, and partially supported through the Bernard W. Gamson Computational Science Center at Yeshiva College. The research of T.J.L. and M.A.A. was partially supported by NSF award no. 1856479. FC acknowledges support from Agence Nationale de la Recherche, grant number ANR-19-CE30-0035-01.

\bibliographystyle{tfo}
\bibliography{refs}

\begin{thebibliography}{82}
\providecommand{\url}[1]{\texttt{#1}}
\providecommand{\urlprefix}{URL }

\bibitem{Stanley_Liquid_2013}
H.E. Stanley, in \emph{Advances in Chemical Physics}, edited by Stuart~A. Rice and Aaron~R. Dinner, Vol. 152  (, , 2013).

\bibitem{Anisimov_Polyamorphism_2018}
M.A. Anisimov, M. Duška, F. Caupin, L.E. Amrhein, A. Rosenbaum and R.J. Sadus,  Phys. Rev. X  \textbf{8}, 011004 (2018).

\bibitem{Tanaka_Liquid_2020}
H. Tanaka,  J. Chem. Phys.  \textbf{153}, 130901 (2020).

\bibitem{Vollhardt_He_1990}
D. Vollhardt and P. Wölfle, \emph{The Superfluid Phases of Helium 3}   (Taylor and Francis, London, UK, 1990).

\bibitem{Schmitt_He_2015}
A. Schmitt, \emph{Introduction to Superfluidity}, \emph{Lecture Notes in Physics}, Vol. 888   (Springer International Publishing, Cham, 2015).

\bibitem{Ohta_H_2015}
K. Ohta, K. Ichimaru, M. Einaga, S. Kawaguchi, K. Shimizu, T. Matsuoka, N. Hirao and Y. Ohishi,  Sci. Rep.  \textbf{5} (16560) (2015).

\bibitem{Zaghoo_H_2016}
M. Zaghoo, A. Salamat and I.F. Silvera,  Phys. Rev. B  \textbf{93}, 155128 (2016).

\bibitem{McWilliams_H_2016}
R.S. McWilliams, D.A. Dalton, M.F. Mahmood and A.F. Goncharov,  Phys. Rev. Lett.  \textbf{116}, 255501 (2016).

\bibitem{Norman_Review_2021}
G.E. Norman and I.M. Saitov,  Phys. Usp.  \textbf{64}, 1094 (2021).

\bibitem{Fried_Hydrogen_2022}
N.R. Fried, T.J. Longo and M.A. Anisimov,  J. Chem. Phys.  \textbf{157}, 101101 (2022).

\bibitem{Henry_Sulfur_2020}
L. Henry, M. Mezouar, G. Garbarino, D. Sifré, G. Weck and F. Datchi,  Nature  \textbf{584}, 382--386 (2020).

\bibitem{Shumovskyi_Sulfur_2022}
N.A. Shumovskyi, T.J. Longo, S.V. Buldyrev and M.A. Anisimov,  Phys. Rev. E  \textbf{106}, 015305 (2022).

\bibitem{Katayama_Phos_2000}
Y. Katayama, T. Mizutani, W. Utsumi, O. Shimomura, M. Yamakata and K. ichi Funakoshi,  Nature  \textbf{403}, 170--173 (2000).

\bibitem{Katayama_Phos_2004}
Y. Katayama, Y. Inamura, T. Mizutani, M. Yamakata, W. Utsumi and O. Shimomura,  Science  \textbf{306} (5697), 848--851 (2004).

\bibitem{Glosli_Liquid_1999}
J.N. Glosli and F.H. Ree,  Phys. Rev. Lett.  \textbf{82}, 4659--4662 (1999).

\bibitem{Sastry_Silicon_2003}
S. Sastry and C.A. Angell,  Nature Mater  \textbf{2}, 739--743 (2003).

\bibitem{Beye_Silicon_2010}
M. Beye, F. Sorgenfrei, W.F. Schlotter, W. Wurth and A. Föhlisch,  Proc. Natl. Acad. Sci.  \textbf{107} (39), 16772--16776 (2010).

\bibitem{Vasisht_Solicon_2011}
V.V. Vasisht, S. Saw and S. Sastry,  Nat. Phys.  \textbf{7}, 549--553 (2011).

\bibitem{Sciorino_Silicon_2011}
F. Sciortino,  Nat. Phys.  \textbf{7}, 523--524 (2011).

\bibitem{Tsuchiya_SeTe_1982}
Y. Tsuchiya and E.F.W. Seymour,  J. Phys. C Solid State Phys.  \textbf{15} (22), L687--L695 (1982).

\bibitem{Brazhkin_SeTe_1999}
V.V. Brazhkin, S.V. Popova and R.N. Voloshin,  Physica B  \textbf{265}, 64--71 (1999).

\bibitem{Cadien_Ce_2013}
A. Cadien, Q.Y. Hu, Y. Meng, Y.Q. Cheng, M.W. Chen, J.F. Shu, H.K. Mao and H.W. Sheng,  Phys. Rev. Lett.  \textbf{110}, 125503 (2013).

\bibitem{Saika_Silica_2000}
I. Saika-Voivod, F. Sciortino and P.H. Poole,  Phys. Rev. E  \textbf{63}, 011202 (2000).

\bibitem{Lascaris_Silica_2014}
E. Lascaris, M. Hemmati, S.V. Buldyrev, H.E. Stanley and C.A. Angell,  J. Chem. Phys.  \textbf{140}, 224502 (2014).

\bibitem{Angell_TwoState_1971}
C.A. Angell,  J. Phys. Chem.  \textbf{75} (24), 3698 (1971).

\bibitem{Angell_Amorphous_2004}
C.A. Angell,  Annual Review of Physical Chemistry  \textbf{55} (1), 559--583 (2004).

\bibitem{Poole_Water_1992}
P.H. Poole, F. Sciortino, U. Essmann and H.E. Stanley,  Nature  \textbf{360}, 324--328 (1992).

\bibitem{Debenedetti_Water_1998}
P.G. Debenedetti,  Nature  \textbf{392}, 127--128 (1998).

\bibitem{Holten_Water_2012}
V. Holten and M.A. Anisimov,  Sci. Rep.  \textbf{2}, 713 (2012).

\bibitem{Holten_Water_2014}
V. Holten, J.C. Palmer, P.H. Poole, P.G. Debenedetti and M.A. Anisimov,  J. Chem. Phys.  \textbf{140}, 104502 (2014).

\bibitem{Gallo_Water_2016}
P. Gallo, K. Amann-Winkel, C.A. Angell, M.A. Anisimov, F. Caupin, C. Chakravarty, E. Lascaris, T. Loerting, A.Z. Panagiotopoulos, J. Russo, J.A. Sellberg, H.E. Stanley, H. Tanaka, C. Vega, L. Xu and L.G.M. Pettersson,  Chem. Rev.  \textbf{116}, 7463--7500 (2016).

\bibitem{Biddle_Water_2017}
J.W. Biddle, R.S. Singh, E.M. Sparano, F. Ricci, M.A. González, C. Valeriani, J.L.F. Abascal, P.G. Debenedetti, M.A. Anisimov and F. Caupin,  J. Chem. Phys.  \textbf{146}, 034502 (2017).

\bibitem{Caupin_Thermodynamics_2019}
F. Caupin and M.A. Anisimov,  J. Chem. Phys.  \textbf{151}, 034503 (2019).

\bibitem{Duska_Water_2020}
M. Duška,  J. Chem. Phys.  \textbf{152}, 174501 (2020).

\bibitem{Holten_Compressibility_2017}
V. Holten, C. Qiu, E. Guillerm, M. Wilke, J. Rička, M. Frenz and F. Caupin,  J. Phys. Chem. Lett.  \textbf{8} (22), 5519--5522 (2017).

\bibitem{Kim_Maxima_2017}
K.H. Kim, A. Späh, H. Pathak, F. Perakis, D. Mariedahl, K. Amann-Winkel, J.A. Sellberg, J.H. Lee, S. Kim, J. Park, K.H. Nam, T. Katayama and A. Nilsson,  Science  \textbf{358} (6370), 1589--1593 (2017).

\bibitem{Pathak_Enhancement_2021}
H. Pathak, A. Späh, N. Esmaeildoost, J.A. Sellberg, K.H. Kim, F. Perakis, K. Amann-Winkel, M. Ladd-Parada, J. Koliyadu, T.J. Lane, C. Yang, H.T. Lemke, A.R. Oggenfuss, P.J.M. Johnson, Y. Deng, S. Zerdane, R. Mankowsky, P. Beaud and A. Nilsson,  Proc. Natl. Acad. Sci.  \textbf{118} (6), e2018379118 (2021).

\bibitem{Hruby_Surface_2014}
J. Hrub\'{y}, V. Vin\v{s}, R. Mare\v{s}, J. Hykl and J. Kalov\'{a},  J. Phys. Chem. Lett.  \textbf{5} (3), 425--428 (2014).

\bibitem{Vins_Surface_2015}
V. Vinš, M. Fransen, J. Hykl and J. Hrubý,  J. Phys. Chem. B  \textbf{119} (17), 5567--5575 (2015).

\bibitem{Vins_Surface_2017}
V. Vinš, J. Hošek, J. Hykl and J. Hrubý,  J. Chem. Eng. Data  \textbf{62} (11), 3823--3832 (2017).

\bibitem{Vins_Surface_2020}
V. Vinš, J. Hykl, J. Hrubý, A. Blahut, D. Celný, M. Čenský and O. Prokopová,  J. Phys. Chem. Lett.  \textbf{11} (11), 4443--4447 (2020).

\bibitem{Gonzalez_Comprehensive_2016}
M.A. Gonz\`alez, C. Valeriani, F. Caupin and J.L.F. Abascal,  J. Chem. Phys.  \textbf{145}, 054505 (2016).

\bibitem{Palmer_Anomalous_2018}
J. Guo, R.S. Singh and J.C. Palmer,  Molecular Physics  \textbf{116} (15-16), 1953--1964 (2018).

\bibitem{Singh_Thermodynamic_2019}
R.S. Singh, J.C. Palmer, A.Z. Panagiotopoulos and P.G. Debenedetti,  J. Chem. Phys.  \textbf{150}, 2224503 (2019).

\bibitem{Debenedetti_Second_2020}
P.G. Debenedetti, F. Sciortino and G.H. Zerze,  Science  \textbf{369} (6501), 289--292 (2020).

\bibitem{Mishima_Decompression_1998}
O. Mishima and H.E. Stanley,  Nature  \textbf{392}, 164--168 (1998).

\bibitem{Kim_WaterExp_2020}
K.H. Kim, K. Amann-Winkel, N. Giovambattista, A. Späh, F. Perakis, H. Pathak, M.L. Parada, C. Yang, D. Mariedahl, T. Eklund, T.J. Lane, S. You, S. Jeong, M. Weston, J.H. Lee, I. Eom, M. Kim, J. Park, S.H. Chun, P.H. Poole and A. Nilsson,  Science  \textbf{370} (6519), 978--982 (2020).

\bibitem{Abascal_General_2005}
J.L.F. Abascal and C. Vega,  J. Chem. Phys.  \textbf{123}, 234505 (2005).

\bibitem{Vega_Simulating_2011}
C. Vega and J.L.F. Abascal,  Phys. Chem. Chem. Phys.  \textbf{13}, 19663--19688 (2011).

\bibitem{Pallares_Anomalies_2014}
G. Pallares, M.E.M. Azouzi, M.A. González, J.L. Aragones, J.L.F. Abascal, C. Valeriani and F. Caupin,  Proc. Natl. Acad. Sci.  \textbf{111} (22), 7936--7941 (2014).

\bibitem{Pallares_Equation_2016}
G. Pallares, M.A. Gonzalez, J.L.F. Abascal, C. Valeriani and F. Caupin,  Phys. Chem. Chem. Phys.  \textbf{18}, 5896--5900 (2016).

\bibitem{Caupin_Minimal_2021}
F. Caupin and M.A. Anisimov,  Phys. Rev. Lett.  \textbf{127}, 185701 (2021).

\bibitem{MFT_PT_2021}
T.J. Longo and M.A. Anisimov,  J. Chem. Phys.  \textbf{156}, 084502 (2022).

\bibitem{Shumovsky2023}
N.A. Shumovskyi and S.V. Buldyrev,  Phys. Rev. E  \textbf{107}, 024140 (2023).

\bibitem{Buldyrev2024}
S.V. Buldyrev,  Cond. Matt. Physics  \textbf{27}, xxxx (2024).

\bibitem{Longo_Interfacial_2023}
T.J. Longo, S.V. Buldyrev, M.A. Anisimov and F. Caupin,  J. Phys. Chem. B  \textbf{127} (13), 3079--3090 (2023).

\bibitem{Sastry_SingularityFree_1996}
S. Sastry, P.G. Debenedetti, F. Sciortino and H.E. Stanley,  Phys. Rev. E  \textbf{53}, 6144--6154 (1996).

\bibitem{Hruby_Twostructure_2004}
J. Hrubý and V. Holten,  Proceedings of the 14th International Conference on the Properties of Water and Steam, Maruzen  pp. 241--246 (2004).

\bibitem{Ciach_Model_2008}
A. Ciach, W. G\'o\ifmmode \acute{z}\else \'{z}\fi{}d\ifmmode~\acute{z}\else \'{z}\fi{} and A. Perera,  Phys. Rev. E  \textbf{78}, 021203 (2008).

\bibitem{kawasaki_diffusion_1966}
K. Kawasaki,  Physical Review  \textbf{145} (1), 224--230 (1966).

\bibitem{glauber_timedependent_1963}
R.J. Glauber,  J. Math. Phys.  \textbf{4} (2), 294--307 (1963).

\bibitem{Shum_Phase_2021}
N.A. Shumovskyi, T.J. Longo, S.V. Buldyrev and M.A. Anisimov,  Phys. Rev. E  \textbf{103}, L060101 (2021).

\bibitem{metropolis_basic_1963}
N. Metropolis and R.L. Ashenhurst,  IEEE Trans. Comput  \textbf{EC-12} (6), 896--904 (1963).

\bibitem{Widom63}
B. Widom,  J. Chem. Phys.  \textbf{39} (6), 2808--2812 (1963).

\bibitem{Ch8_Hassan_Mesothermo_2010}
H. Behnejad, J.V. Sengers and M.A. Anisimov, in \emph{Applied Thermodynamics of Fluids}, edited by Anthony~R. Goodwin, Cor~J Peters and Jan Sengers, Chap.~10  (Royal Society of Chemistry, Cambridge, 2010), pp. 321--367.

\bibitem{Wang_isomorphism_2008}
J. Wang, C.A. Cerdeiri\~na, M.A. Anisimov and J.V. Sengers,  Phys. Rev. E  \textbf{77}, 031127 (2008).

\bibitem{Fisher_finite_1972}
M.E. Fisher and M.N. Barber,  Phys. Rev. Lett.  \textbf{28}, 1516--1519 (1972).

\bibitem{Hansen_Liquid_2013}
J.P. Hansen and I.R. McDonald, \emph{Theory of Simple Liquids: with Applications to Soft Matter}, 4th ed.   (Academic Press, Oxford, 2013).

\bibitem{Kim_Crossover_2003}
Y.C. Kim, M.A. Anisimov, J.V. Sengers and E. Luijten,  J. Stat. Phys.  \textbf{110}, 591--609 (2003).

\bibitem{Ch8_Kostko_Crossover_2005}
M.A. Anisimov, A.F. Kostko, J.V. Sengers and I.K. Yudin,  J. Chem. Phys.  \textbf{123}, 164901 (2005).

\bibitem{Fisher_scattering_1967}
M.E. Fisher and R.J. Burford,  Phys. Rev.  \textbf{156}, 583--622 (1967).

\bibitem{Binder_Calculation_1982}
K. Binder,  Phys. Rev. A  \textbf{25}, 1699--1709 (1982).

\bibitem{Debenedetti_metastable_1996}
P.G. Debenedetti, \emph{Metastable liquids}   (Princeton University Press, Princeton, NJ, 1996).

\bibitem{Arvengas2011}
A. Arvengas, E. Herbert, S. Cersoy, K. Davitta and F. Caupin,  J. Phys. Chem  \textbf{115}, 14240–14245 (2011).

\bibitem{Lu_second_2006}
Y.J. Lü and B. Wei,  Appl. Phys. Lett.  \textbf{89} (16), 164106 (2006).

\bibitem{Rogers_evidence_2016}
T.R. Rogers, K.Y. Leong and F. Wang,  Sci. Rep.  \textbf{6}, 33284 (2016).

\bibitem{Wang_second_2019}
X. Wang, K. Binder, C. Chen, T. Koop, U. Pöschl, H. Su and Y. Cheng,  Phys. Chem. Chem. Phys.  \textbf{21}, 3360--3369 (2019).

\bibitem{Malek_surface_2019}
S.M.A. Malek, P.H. Poole and I. Saika-Voivod,  J. Chem. Phys.  \textbf{150}, 234507 (2019).

\bibitem{Gorfer_high_2023}
A. Gorfer, C. Dellago and M. Sega,  J. Chem. Phys.  \textbf{158}, 054503 (2023).

\bibitem{Hrahsheh_second_2024}
F. Hrahsheh, I. Jum’h and G. Wilemski,  J. Chem. Phys.  \textbf{160}, 114504 (2024).

\bibitem{Widom_surface_1965}
B. Widom,  J. Chem. Phys.  \textbf{43}, 3892--3897 (1965).

\bibitem{Mauro_Pressure_2020}
M. Sellitto,  J. Chem. Phys.  \textbf{153}, 161101 (2020).

\end{thebibliography}

\appendix

\section{Sellitto's ``Ghost Site'' Method for Pressure Calculations}\label{Appendix}
In addition to the traditional Widom-insertion method to calculate the pressure in a MC system (see Section~\ref{Sec_WidomMethod}), we also considered a contemporary method to calculate the pressure. In this appendix, we describe this alternative approach and provide initial pressure calculations for the blinking-checkers model. 

Sellitto's ghost-site method, described in Ref.~\cite{Mauro_Pressure_2020}, introduces $n_\lambda=100$ ghost sites, each identified as a specific $\lambda$. The ghost sites exchange states with randomly selected sites in the lattice. The ghost sites can be empty or occupied by particles of types 1 and 2. A particle of type 1 or 2 in the $i$th $(i=1,2,\ldots,n_\lambda)$ ghost site has a potential $k_\text{B} T \ln\left(i/n_\lambda\right)$. If the ghost site is empty, it has zero potential energy. 

\begin{figure}[b!]
    \centering
    \includegraphics[width=0.49\linewidth]{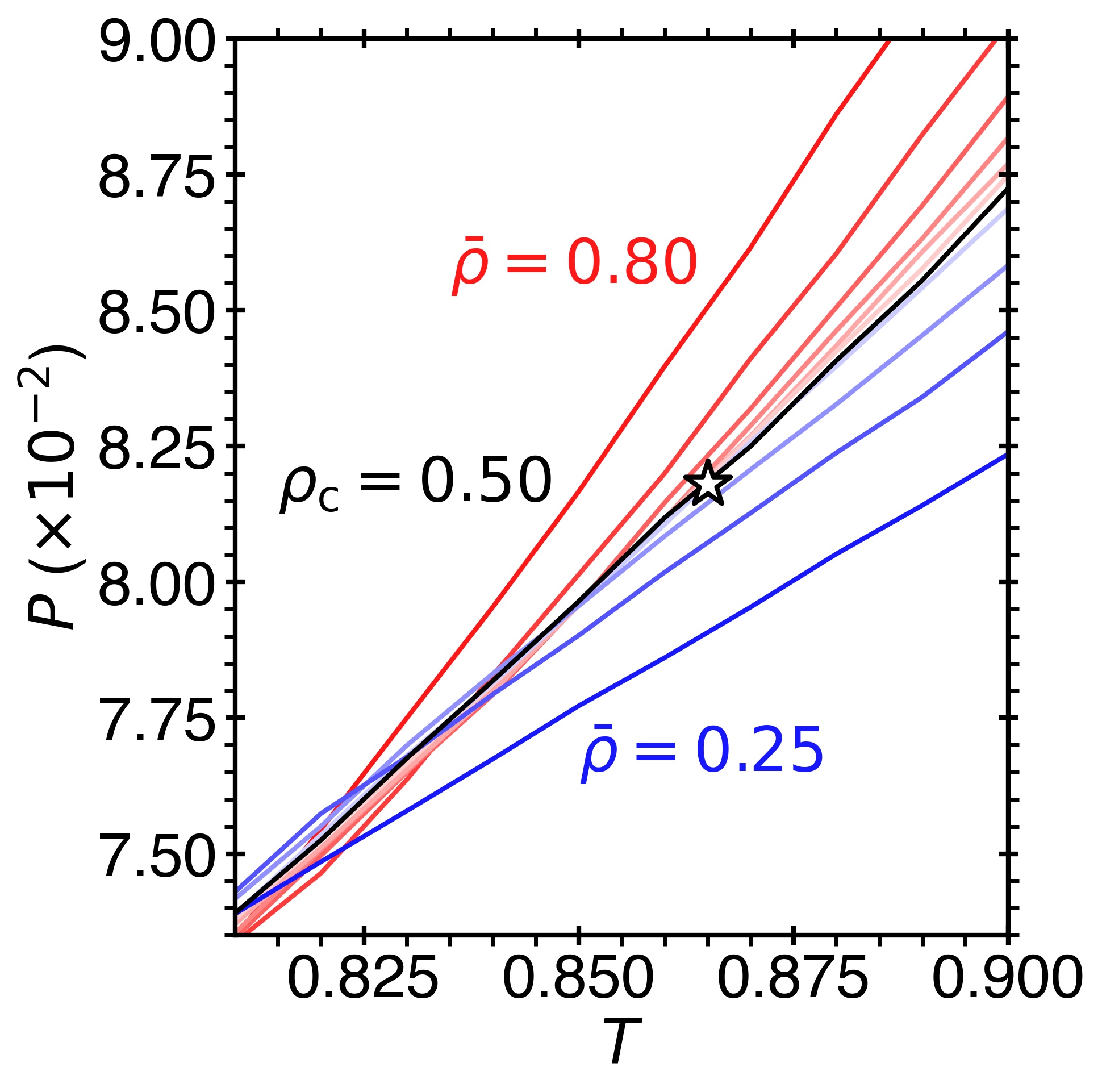}
    \caption{Pressure-density isochores calculated with use of Sellitto's ghost-site method in the vicinity of the LGCP. Isochores are given in steps of the coexisting density of the two liquid phases $\Delta\rho = 0.05$ in which those with blue shades are below $\rho_\text{c}$, black is the critical isochore ($\rho_\text{c}$), and those with red shades are above $\rho_\text{c}$. The critical point (star) is approximately determined from the highest temperature and pressure of the intersecting isochores. While the prediction of the critical temperature ($T_\text{c} = 0.865$) and critical density ($\rho_\text{c} = 0.50$) are in good agreement with the prediction from the Widom-insertion method (see Fig.~\ref{Fig_Phase_Diagrams}a), the prediction for the critical pressure ($P_\text{c} = 8.18\times 10^{-2}$) does not share the same order of magnitude as that from the traditional method nor from the meanfield prediction~\cite{Longo_Interfacial_2023}.}
    \label{Fig_Ghost_PT}
\end{figure}

When a MC time step becomes equal to a multiple of $n_p=32$, each ghost site is allowed to exchange its content with a randomly selected site in the lattice. If a particle of type $k$ is moved to a ghost site, it has a potential energy $U_{\lambda,k=1,2}=-k_\text{B}T\ln\left(i/n_\lambda\right)$ or $U_{\lambda,k=0}=0$. Alternatively, if a particle of type $k$ is moved from the ghost site to a randomly selected lattice site, $i$, it contributes to the total internal energy, $U$, through the introduction of the term
\begin{equation}
    U_{i,k}=\sum_{j\in\mathcal{L}_i} \epsilon(s_k,s_i)
\end{equation}
For instance, suppose that site $i$ is occupied by a particle of type $k$ and the ghost site is occupied by a particle of type $l$. Accordingly, the probability of an exchange of particles between the ghost site $\lambda$ and the lattice site $i$ is computed as $\exp(-\Delta U/k_\text{B}T)$, where  $\Delta U= U_{\lambda,l}+U_{i,k} -U_{\lambda,k}-U_{i,l}$. 

The average occupancy of a ghost site $i$ is $N(\lambda_i)\leq 1$, such that the pressure is computed as
\begin{equation}\label{eq:Plambda}
    P=\rho k_\text{B}T\sum_{i=1}^{n_\lambda} \frac{N(\lambda_i)}{i}
\end{equation}
Note that for non-interacting particles of density $\rho$, this method gives the exact result, $P=-k_\text{B}T\ln(1-\rho)$.

The pressure-temperature phase diagram in the vicinity of the LGCP, calculated with use of Sellitto's method, is presented in Fig.~\ref{Fig_Ghost_PT}. While the prediction of the critical temperature ($T_\text{c} = 0.865$) and critical density ($\rho_\text{c} = 0.50$) are in good agreement with the prediction from the Widom-insertion method (see Fig.~\ref{Fig_Phase_Diagrams}a), the prediction for the critical pressure ($P_\text{c} = 8.18\times 10^{-2}$) does not share the same order of magnitude as that from the traditional method nor from the meanfield prediction~\cite{Longo_Interfacial_2023}. Consequently, in the main text, all pressure calculations were performed with use of the Widom-insertion method. That being said, Sellitto's method is much more efficient computationally than the Widom-insertion method, since, in the traditional method, the pressure calculation must be repeated for all densities from $\rho = 0$ to the state point (see Eq.~(\ref{Eq_Pressure_Widom}) and the discussion in Section~\ref{Sec_WidomMethod}). Therefore, Sellitto's method was found to be useful for quickly generating and evaluating the phase diagram for a given set of model parameters ($\omega_{11}$, $\omega_{22}$, $\omega_{12}$, $\tilde{e}$, and $\tilde{s}$).

\end{document}